\documentclass{jfm}
\usepackage{graphicx,floatrow}
\usepackage{dcolumn}
\usepackage{bm}

\usepackage{amsmath}
\DeclareMathOperator\arctanh{arctanh}

\newcommand{\beq}{\begin{equation}}
\newcommand{\eeq}{  \end{equation}}
\newcommand{\beqa}{\begin{eqnarray}}
\newcommand{\eeqa}{  \end{eqnarray}}

\newcommand{\grad}{{\boldsymbol \nabla}}

\def\strutdepth{\dp\strutbox}
\def\nw#1{\strut\vadjust{\kern-\strutdepth\vtop to0pt{\vss\hbox to\hsize
{\hskip\hsize\hskip5pt$\leftarrow$\hss\strut}}}{\em #1}}

\begin{document}
\title{ Self-similar breakup of polymeric threads as described by
the Oldroyd-B model}

\author{
J. Eggers$^1$, M. A. Herrada$^2$, and J. H. Snoeijer$^3$
}

\affiliation{
	$^1$School of Mathematics,
	University of Bristol, University Walk,
	Bristol BS8 1TW, United Kingdom  \\
        $^2$E.S.I., Universidad de Sevilla, Camino de los Descubrimientos
          s/n 41092, Spain \\
	$^3$Physics of Fluids Group, Faculty of Science and Technology,
	Mesa+ Institute, University of Twente, 7500 AE Enschede,
        The Netherlands
    }

\maketitle

\begin{abstract}
When a drop of fluid containing long, flexible polymers breaks up,
it forms threads of almost constant thickness, whose size
decreases exponentially in time. Using an Oldroyd-B fluid as a model,
we show that the thread profile, rescaled by the thread thickness,
converges to a similarity solution. Using the correspondence between
viscoelastic fluids and non-linear elasticity, we derive similarity
equations for the full three-dimensional axisymmetric flow field in
the limit that the viscosity of the solvent fluid can be neglected.
A conservation law balancing pressure and elastic energy
permits to calculate the thread thickness exactly. The explicit form of the
velocity and stress fields can be deduced from a solution of the
similarity equations. Results are validated by detailed comparison with
numerical simulations.
\end{abstract}

\section{Introduction}
\begin{figure}
\centering
\includegraphics[width=0.96\hsize]{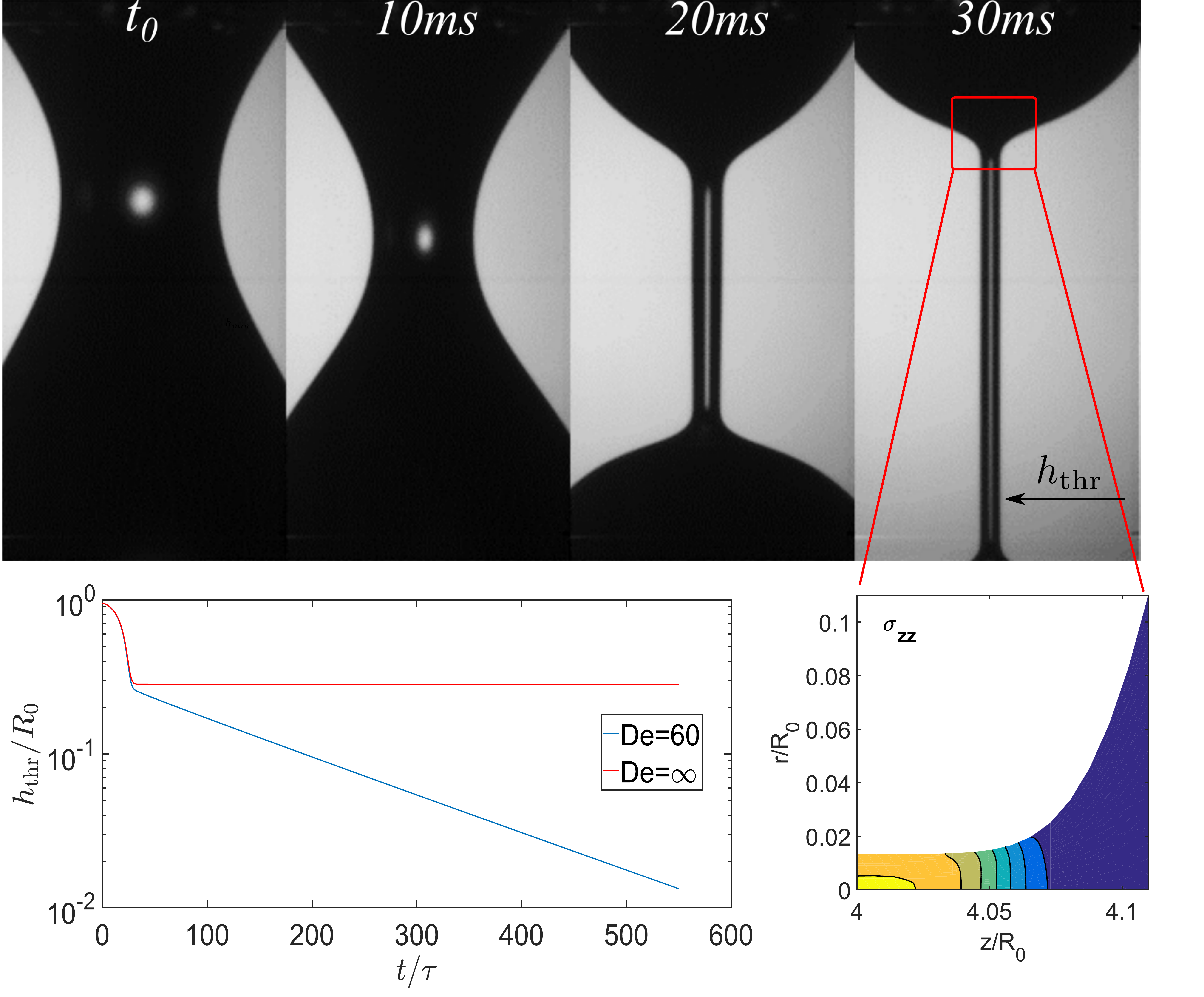}
\caption{\label{bridge_exp}
An illustration of the problem studied in this paper. On the top,
a sequence of experimental images showing the pinch-off of a drop of
polymer solution (PAM) between two solid plates, forming a uniform thread
of radius $h_{\rm thr}(t)$ \cite[]{DVB18}. On the bottom, our numerical
simulation of the Oldroyd B model for flexible polymer solutions,
with dimensionless parameters (defined in Section~\ref{sub:sim}),
chosen as in \cite{TLEAD18}: ${\rm De} = \lambda/\tau = 60$, ${\rm Oh} = 3.16$,
$S = \eta_s/\eta_0 = 0.25$. The elasto-capillary number is
$G = \eta_p R_0/(\lambda\gamma) = 0.0395$. Times are made dimensionless with
the capillary time scale
$\tau = \sqrt{\rho R_0^3/\gamma}$, and lengths are given in
units of the initial bridge radius $R_0$.
On the left, $h_{\rm thr}(t)$ is shown to decrease exponentially in
time, with decay exponent $1/3\lambda$, where $\lambda$ is the polymer
timescale (blue line). For comparison, we also include a simulation
with infinite relaxation time, keeping $G$ constant.
On the right, a closeup of the corner region between the
thread and the drop, showing the axial stress $\sigma_{zz}$. The
stress is very high inside the thread, and decays very rapidly in the
transition region.
}
\end{figure}

When a drop of water falls from a tap, it will break into two or more
pieces under the action of surface tension \cite[]{EV08}. Similarly,
holding a drop between two solid plates which are then separated to make
this liquid bridge unstable, breakup is observed at some finite time
$t_0$. Near breakup time, drop profiles and the velocity field
are characterised by similarity solutions \cite[]{E93,EF_book}, which
describe the evolution toward smaller and smaller scales.

If even a small
amount of long, flexible polymer is added, the singularity is inhibited
\cite[]{BVER81,ABMK01,CDK99,CEFLM06,MH10,EF_book}, and a thread of almost
uniform thickness
is formed, as shown in Fig.~\ref{bridge_exp}. This is because polymers
are stretched in the extensional flow near breakup, but resist the
stretching, producing an axial stress, sometimes quantified as an
extensional viscosity. This slows down the pinching, and
results in a uniform thread thickness. In the framework of the
Oldroyd-B model for polymer solutions, the characteristic relaxation
time $\lambda$ of the polymer selects a constant extension rate
$\dot{\epsilon}$ inside the thread, which leads to an exponential
thinning
\beq
h_{\rm thr} = h_0 \exp\left(-\frac{t}{3\lambda}\right)
\label{hmin}
\eeq
of the thread radius \cite[]{BER90,CEFLM06,EV08}. This is
well confirmed by experiments \cite[]{BER90,AM01,ABMK01,CEFLM06}
in both high and low viscosity solvents, as well as in full numerical
simulations of the Oldroyd-B equations 
\cite[]{EHL06,BAHPMB10,TLEAD18}.

In Fig.~\ref{bridge_exp} (red curve on bottom left) we also show
the minimum thread radius for a simulation in which we have taken the
limit of infinite relaxation time. As a result,
elastic stresses do not relax, and a stationary minimum thread thickness
is reached at a point where elastic and surface tension forces balance.
Below we will show that the time-dependent problem of an exponentially
thinning thread is closely related to the stationary elastic problem.
Both are described by the same similarity solution, respectively,
in the limit of long times, and of vanishing elasto-capillary number
(given that contributions from the solvent are small).

One of the main outstanding questions is to calculate the prefactor
$h_0$ theoretically, and to relate it to the extensional viscosity
inside the thread. By measuring the thread thickness one could then
infer the viscosity, and thus use the device as a rheometer \cite[]{BER90}.
However, it was pointed out by \cite{CEFLM06} that a proper description
of the thread cannot result from a force balance within the uniform thread
alone, since the driving force of the motion results from the capillary
pressure {\it difference} between the thread and the drops which connect
to it (cf. sequence at the top of Fig.~\ref{bridge_exp}). To calculate the
resistive force, one has to consider the flow of liquid emptying from the
thread into the drops (of which in Fig.~\ref{bridge_exp} there is one
on either end of the thread). As a result, the task is to understand
the transition region connecting the thread and the drop, which has the
form of a similarity solution \cite[]{CEFLM06,TLEAD18}: if both the
axial and radial coordinates are rescaled by the thread thickness
$h_{\rm thr}$, the drop profile in the corner region connecting the thread
and the drop falls onto a universal curve.

In \cite{CEFLM06} the similarity solution for the corner was calculated
in the framework of lubrication theory, which is valid if the interface
slope is assumed small, so the flow inside the thread can be described
within a one-dimensional model. Clearly, this is not satisfied throughout
the corner as seen in Fig.~\ref{bridge_exp}; we will in fact find that
the slope of the similarity solution grows exponentially away from the
thread. It is also clear that the stress distribution has a significant
radial dependence (cf. the bottom right of Fig.~\ref{bridge_exp}),
which cannot be captured by a one-dimensional model.

A more recent numerical study \cite[]{TLEAD18} has confirmed
that the corner solution is fully three-dimensional axisymmetric,
while it has the similarity structure anticipated in \cite{CEFLM06}.
The present paper closes this gap by calculating the full self-similar
structure of the three-dimensional axisymmetric equations, both numerically
and theoretically. Following many previous papers, we do this in the
framework of the Oldroyd-B model, which has become a standard model
for the description of solutions of long, flexible polymers
\cite[]{BAH87,L99,MS15}, including capillary flows \cite[]{EV08}.
The stress contains contributions from both the polymers themselves,
and a Newtonian contribution coming from the solvent.
Its main simplifying features are that polymers are modelled with a
single characteristic relaxation time of the polymer $\lambda$, and that
each polymer strand is assumed infinitely extensible, so that
the extensional viscosity can become arbitrarily large.

As a result, the exponential thinning regime described by \eqref{hmin} will
continue forever in the Oldroyd-B model, which allows us to focus on
this critical part of the pinching dynamics. If finite extensibility
of the polymer is taken into account \cite[]{FL04,WABE05}, as in the
so-called FENE-P model \cite[]{BAH87}, pinching
proceeds in a localised fashion, similar to the Newtonian case
\cite[]{Ren02,FL04}. However, this mode of breakup does not appear to be
a realistic description of experiment, as instead
\cite[]{OM05,SWE08,SGEW12} the polymer thread
undergoes a spatially periodic instability, in the course of which the
highly stressed thread partially relaxes, to form a series of small
droplets. This has been called the blistering instability
\cite[]{SWE08,SGEW12}, whose theoretical description \cite[]{E_pol14}
has recently been confirmed experimentally \cite[]{DVB18}. 

In this paper, we give the first consistent analysis of polymeric
pinching using the full three-dimensional, axisymmetric Oldroyd-B equations.
In the next section, we present the Oldroyd-B equations of motion. We
use a powerful new code based on mapping the physical domain unto
a simple rectangular domain \cite[]{Herrada2016,DHHVRKEB18}, together
with a logarithmic transformation of the polymeric stress
tensor \cite[]{TLEAD18}, in order to study the self-similar structure
of the polymeric pinching problem in detail. We find that contrary
to previous assumptions, owing to the stress contribution from the
solvent, the stress distribution inside the thread has a non-constant
radial distribution. This suggests the similarity solution in general is
non-universal, and dependent on boundary conditions. However, the contribution
from the solvent is often small for realistic parameter values, as
measured by a dimensionless parameter $\overline{v}_0$ which we identify
below. As a result, the main focus of our paper is on calculating
the universal similarity solution in the limit of vanishing solvent
contribution; the non-universal part comes from the solvent contribution. 

In Section~\ref{sec:sim} we present the similarity equations describing
the corner region between the polymer thread and the drop including 
solvent contributions. Since inertia is found to be subdominant asymptotically,
the tension in each cross section is conserved. To solve the similarity
equations for vanishing solvent parameter $\overline{v}_0$, in
Section~\ref{sec:elastic} we consider a related problem in non-linear
elasticity: the collapse
of a cylinder of elastic material under surface tension. As the elastic
bridge deforms, elastic stresses build up and eventually balance surface
tension to establish a new stationary equilibrium state. For small
elastic modulus $\mu$, a thread of almost uniform thickness is formed,
whose radius goes to zero in the limit of vanishing $\mu$. We show that
this limit is described by a similarity solution, which is {\it identical}
to the similarity solution for the time-dependent collapse of a
viscoelastic fluid bridge.

Using Lagrangian coordinates with respect to the reference state of
a cylindrical elastic bridge, we derive a simplified set of similarity
equations for the elastic problem in the limit $\mu\rightarrow 0$.
The elastic similarity equations are shown to obey a novel conservation
law involving the pressure and the elastic energy, which is closely
related to Eshelby's elastic energy-momentum tensor \cite[]{E75}.
Physically, it represents invariance of the equations under a relabelling
of Lagrangian coordinates.

With the help of this conservation law, we are able to calculate
the thickness of the elastic thread, without having to solve the
full three-dimensional similarity equations. Remarkably, the result
agrees with what was found by us earlier \cite{CEFLM06} using a
one-dimensional description, because the description correctly
respects the conservation law. However, the spatial structure of
the similarity solution is different from what the earlier one-dimensional
description predicts. We then solve the similarity equations
numerically to yield the self-similar interface profile and the
stress distribution in the interior. This then solves the problem both of the
time-dependent pinching of a viscoelastic thread, and the collapse of
an elastic thread.

In a final discussion, we compare numerical results to our similarity
theory, and outline future directions of research.

\section{Equations of motion and numerical simulations}
\label{sec:simulation}
The Oldroyd-B model can be described by the following set of
equations \cite[]{BAH87,MS15}:
\beqa
\label{equ1}
&& \rho\left(\frac{\partial {\bf v}}{\partial t} +
\left({\bf v}\cdot\grad\right){\bf v} \right)
= -\grad p + \grad\cdot\boldsymbol{\sigma}_p
+ \eta_s\triangle {\bf v}, \\
\label{equ2}
&& \frac{\partial \boldsymbol{\sigma}_{p}}{\partial t} +
\left({\bf v}\cdot\grad\right)
\boldsymbol{\sigma}_p =
(\grad{\bf v})^T\cdot\boldsymbol{\sigma}_p +
\boldsymbol{\sigma}_p\cdot(\grad{\bf v}) -
\frac{\boldsymbol{\sigma}_p}{\lambda} +
\frac{\eta_p}{\lambda}\dot{\boldsymbol{\gamma}} \\
\label{equ3}
&& \grad\cdot{\bf v} = 0,
\eeqa
where $\dot{\boldsymbol{\gamma}} =
(\grad {\bf v}) + (\grad {\bf v})^T$ is the
rate-of-deformation tensor, and $\rho$ the fluid density.
As seen on the right of the momentum
equation \eqref{equ1}, the total stress on a fluid element is the
sum of two contributions, coming from the solvent and from the polymer.
The solvent contributes a Newtonian stress, where $\eta_s$ the dynamic
viscosity of the solvent, while $\boldsymbol{\sigma}_p$ is the deviatoric
part of the polymeric stress.

The Oldroyd-B model can be derived as the continuum version of a
solution of non-interacting model polymers, each of which consists of
two beads, experiencing Stokes' drag, and connected by a Hookean
spring. Thus on one hand, springs become stretched when beads move
apart as described by the flow. On the other hand, the resulting tension
in the string contributes to the polymeric stress $\boldsymbol{\sigma}_p$,
described by the equation of motion \eqref{equ2}; \eqref{equ3} enshrines
incompressibility of the flow.

In the limit of small shear rates (so that polymers are hardly stretched
at all), \eqref{equ1}-\eqref{equ3} describe a Newtonian fluid of
total dynamic viscosity $\eta_0 = \eta_s + \eta_p$; thus $\eta_p$ is
known as the polymeric contribution to the viscosity.
Polymeric stress
relaxes at a rate $\lambda$, as seen from the second to last term on the
right of \eqref{equ2}, while stretching by the flow is described by the first
two terms on the right. In particular, in the case of an extensional
flow with constant extension rate $\dot{\epsilon}$, the polymeric
stress will grow exponentially if $\dot{\epsilon}\lambda > 1/2$.
This follows from a comparison of the stretching and relaxation terms.
We will see below that in fact $\dot{\epsilon}\lambda = 2/3$ inside the thread.

From a continuum perspective, the first four terms of \eqref{equ2} are
known as the ``upper convected derivative''
\beq
\overset{\triangledown}{\boldsymbol{\sigma}}_p =
\frac{\partial{\boldsymbol{\sigma}}_p}{\partial t} +
\left({\bf v}\cdot\grad\right){\boldsymbol{\sigma}}_p
- \left(\grad{\bf v}\right)^T{\boldsymbol{\sigma}}_p -
{\boldsymbol{\sigma}}_p\left(\grad{\bf v}\right).
\label{ucd}
\eeq
Its form can be derived purely from the requirement that
the polymeric stress ought to transform consistently as a
covariant tensor \cite[]{MS15}.

On account of the axisymmetry of the problem, the velocity
field can be written
${\bf v} = v_r{\bf e}_r + v_z{\bf e}_z$ in cylindrical coordinates,
and the polymeric stress tensor is
$\boldsymbol{\sigma}_p = \sigma_{rr}{\bf e}_r\otimes{\bf e}_r +
\sigma_{rz}{\bf e}_r\otimes{\bf e}_z + \sigma_{zz}{\bf e}_z\otimes{\bf e}_z
+ \sigma_{\theta\theta}{\bf e}_\theta\otimes{\bf e}_\theta$.
The stress boundary condition at the free surface is
\beq
{\bf n}\cdot(\boldsymbol{\sigma}_p + \eta_s\dot{\boldsymbol{\gamma}}) =
(p-\gamma\kappa){\bf n},
\label{stress_bc}
\eeq
where
\beq
\kappa = \frac{1}{h(1+h_{z}^2)^{1/2}} -
\frac{h_{zz}}{(1+h_{z}^{2})^{\frac{3}{2}}}, \quad
{\bf n} = \frac{{\bf e}_r-{\bf e}_z h_z}{(1+h_{z}^2)^{1/2}}
\label{kappa}
\eeq
are (twice) the mean curvature and the surface normal, respectively. The
coefficient $\gamma$ is the surface tension.
If $h(z,t)$ is the thread profile, the kinematic boundary condition becomes
\beq
\frac{\partial h}{\partial t} + v_z(z,h)\frac{\partial h}{\partial z} =
v_r(z,h).
\label{kin_bc}
\eeq

The numerical problem of solving  \eqref{equ1}-\eqref{equ3} with
boundary conditions \eqref{stress_bc} and \eqref{kin_bc} under
conditions of breakup is highly demanding, for a number of reasons.
First, the Oldroyd-B equation is well known to exhibit numerical
instabilities (for reasons which are not well understood) if
$\dot{\epsilon}\lambda$ is of order unity \cite[]{Fattal2004}. This is
sometimes known as the high Weissenberg number problem (HWNP).
Second, it is crucial to
correctly describe the balance between surface tension forces and
the force exerted by the polymer, both of which go to zero in the limit
of vanishing thread thickness.

\subsection{Logarithmic transformation}
\label{sub:log}
In order to avoid the HWNP as $\dot{\epsilon}\lambda$ is of order unity,
a partial (2D) matrix-logarithm transformation of the polymeric stress
tensor $\boldsymbol{\sigma}_p$ is applied in this work
\cite[]{Fattal2004,TLEAD18}. The idea behind these transformations is
to replace \eqref{equ2}, which allows to advance the stresses
$\sigma_{rr}$, $\sigma_{zz}$ and  $\sigma_{rz}$ in time, with three
equations for the elements of a 2x2  symmetric positive-definite
matrix $\boldsymbol{\psi}$. The first step is to decompose this matrix as
\beq
\boldsymbol{\psi}=\left(
\begin{array}{ccc}
\psi_{zz} &  \psi_{rz}  \\
\psi_{rz} &  \psi_{rr}
\end{array}
\right) =\mathbf{ R}\left(
\begin{array}{cc}
\ln(\Lambda_1) & 0 \\
0 & \ln(\Lambda_2)  \\
\end{array}
\right) \mathbf{ R}^T, \label{equaNumeric1}
\eeq
$\ln(\Lambda_1)$ and $\ln(\Lambda_2)$ being the eigenvalues of
$\boldsymbol{\psi}$, and the columns of $\mathbf{R}$ containing their
normalised eigenvectors: $\mathbf{R}\mathbf{R}^T=\mathbf{I}$.
The second step is to construct the conformation tensor $\mathbf{A}$:
\beq
\mathbf{A}=\left(
    \begin{array}{cc}
      A_{zz}& A_{rz}\\
      A_{rz} & A_{rr}\\
    \end{array}
  \right)=
\mathbf{ R}\left(
\begin{array}{cc}
\Lambda_1 & 0 \\
0 & \Lambda_2  \\
\end{array}\right) \mathbf{ R}^T.
\label{equaNumeric2}
\eeq

Then, the stresses $\sigma_{zz}$, $\sigma_{rz}$ and $\sigma_{rr}$ can be
expressed as function of the elements of $\boldsymbol{\psi}$ with
the help of $\mathbf{A}$:
\beq
\sigma_{zz} = \eta_p(A_{zz}-1)/\lambda,\quad \sigma_{rr} =
\eta_p(A_{rr}-1)/\lambda,\quad \sigma_{rz} = \eta_pA_{rz}/\lambda.
\label{equaNumeric3}
\eeq
Finally, the equation of motion for $\boldsymbol{\psi}$ is
\beq
\frac{\partial \boldsymbol{\psi}}{\partial t}+(\boldsymbol {v}
\cdot \grad)\boldsymbol{\psi}-
(\boldsymbol{\Omega}\boldsymbol{\psi}-\boldsymbol{\psi}\boldsymbol{\Omega})-
2 \mathbf{B}=\frac{1}{\lambda}\left[\boldsymbol{A}^{-1}-\mathbf{I}\right],
\label{equaNumeric4}
\eeq
with matrices $\boldsymbol{\Omega}$, $\boldsymbol{B}$  given by
\beq
 \boldsymbol{\Omega}=\mathbf{ R}\left(
  \begin{array}{cc}
  0& \omega \\
   -\omega & 0 \\
   \end{array}
   \right)\mathbf{ R}^T,\quad  \mathbf{B}=\mathbf{ R}\left(
  \begin{array}{cc}
  M_{11}& 0 \\
   0 & M_{22} \\
   \end{array}
  \right)\mathbf{ R}^T.
\label{equaNumeric5}
\eeq
Here
\beq
\mathbf{M}=\left(
    \begin{array}{cc}
      M_{11}& M_{12} \\
      M_{12} & M_{22} \\
    \end{array}
  \right)=
  \mathbf{ R}^T(\grad \mathbf{v})_{2d}^T\mathbf{ R},\quad
  (\grad \mathbf{v})_{2d}=\left(
    \begin{array}{cc}
      \frac{\partial v_z}{\partial z}&  \frac{\partial v_r}{\partial z} \\
     \frac{\partial v_z}{\partial r} & \frac{\partial v_r}{\partial r}\\
    \end{array}
    \right),
\label{equaNumeric6}
\eeq
and $\omega=[\Lambda_2M_{12}+M_{21}\Lambda_1]/[\Lambda_2-\Lambda_1]$.
In case that $\Lambda_1=\Lambda_2$, matrices
$\boldsymbol{\Omega}$, $\boldsymbol{B}$  are replaced by
\beq
\boldsymbol{\Omega}=\mathbf{0},\quad \mathbf{B}=
\frac{1}{2}[(\mathbf{\grad v})_{2d}^T+(\mathbf {\grad v})_{2d}].
\label{equaNumeric7}
\eeq

\subsection{Mapping technique}
\label{sub:map}
The numerical technique used in this study is a variation of that
developed in \cite{Herrada2016}. The spatial physical domain occupied
by the fluid is mapped onto a rectangular domain
$0\leq \eta\leq 1$,  $0\leq \xi\leq L$ by means of the coordinate
transformation $\eta=r/h(z,t)$ and $\xi=z$. Then each variable
($v_z$, $v_r$, $p$, $\psi_{rr}$, $\psi_{rz}$, $\psi_{zz}$,
$\sigma_{\theta\theta}$ and  $h$) and all its spatial and temporal
derivatives, which appear in the transformed equations, are written
as a single symbolic vector. For example, seven symbolic elements are
associated to the axial velocity, $v_z$: $v_z$,
$\frac{\partial v_z}{\partial \eta}$,
$\frac{\partial v_z}{\partial \xi}$,
$\frac{\partial^2 v_z}{\partial^2 \eta}$,
$\frac{\partial^2 v_z}{\partial^2 \xi}$,
$\frac{\partial^2 v_z}{\partial\eta\xi}$, and
$\frac{\partial v_z}{\partial \tau}$, $\tau=t$ being the
transformation for time. The next step is to use a symbolic
toolbox to calculate the analytical Jacobians of all the equations
with respect to  the symbolic vector. Using these analytical
Jacobians, we generate functions which can be evaluate later point
by point once the transformed domain is discretized in space
and time. In this work, we used the MATLAB tool matlabFunction to
convert the symbolic Jacobians  and equations in  MATLAB functions.

Next we carry out the temporal and spatial discretization of the
transformed domains. The spatial domain is discretized using a set
of $n_\xi=850$  equally spaced collocation points in the axial ($\xi$)
direction, while a set of $n_\eta=20$ of equally spaced collocation
points are used in the radial ($\eta$) direction. Fourth-order finite
differences are employed to compute the collocation matrices associated
with the discretized points. In the time domain, second-order backward
finite differences are used for the discretization.

The final step is to solve at each time step the non-linear system of
discretized equations using a Newton procedure, where the numerical
Jacobian is constructed using the spatial collocations matrices and
the saved analytical functions. Since the method is fully implicit,
a relatively large  fixed time step, $dt/\tau=0.1$, was used in the simulation.

\subsection{Numerical results and parameters}
\label{sub:sim}
As a representative example, we simulate a liquid bridge with periodic
boundary conditions, using the parameters of \cite{TLEAD18}. The initial
condition is a cylinder of radius $R_0$, with a sinusoidal perturbation
of small amplitude $\epsilon$ added to it:
\beq
\frac{h(z,0)}{R_0} = 1 - \epsilon\cos\left(\frac{z}{2 R_0}\right),
\quad -2\pi \le \frac{z}{R_0}\le 2\pi,
\label{init}
\eeq
and $\epsilon = 0.05$. Only half of the domain was simulated, using
symmetry.

We define the dimensionless parameters of the simulation as in
\cite{CEFLM06}. First, the Deborah number ${\rm De} = \lambda/\tau$
is the dimensionless relaxation time, compared to the capillary
time scale $\tau = \sqrt{\rho R_0^3/\gamma}$. The Ohnesorge number
${\displaystyle {\rm Oh} = \eta_0/\sqrt{R_0\rho\gamma}}$ measures the
relative importance of viscous and inertial effects, while
$S = \eta_s/\eta_0$ is the solvent viscosity ratio. The ratio
$\mu = \eta_p / \lambda$ defines an elastic constant; it measures the
effective shear modulus of the viscoelastic fluid on time scales
shorter than the relaxation time $\lambda$. Its dimensionless version
\beq
G = \mu R_0/\gamma = \eta_p R_0 / (\lambda\gamma)
\label{G}
\eeq
is known as the elasto-capillary number.

In most of the plots shown below, we choose the same parameter values
as those in \cite{TLEAD18}:
${\rm De} = 60$, ${\rm Oh} = 3.16$, $S = \eta_s/\eta_0 = 0.25$.
The elasto-capillary number is $G = 0.0395$. Our mapping technique
allows for the description of all fields with high accuracy, avoiding
the singular behaviour of the stress near the free surface reported
previously \cite[]{TLEAD18}. The use of the logarithmic transformation
described in Section~\ref{sub:log}, together with a fully implicit
formulation, allows for superior stability. As a result, we were able
to follow the dynamical evolution to well beyond a dimensionless
time of $t/\tau = 500$. This corresponds to the minimum thread radius
being smaller by a factor of 1/4 than in \cite{TLEAD18}. 

\section{Similarity description}
\label{sec:sim}

\subsection{Solution inside the thread}
As shown in \cite{CEFLM06}, and illustrated in Fig.~\ref{bridge_exp},
there are three different regions involved
in this problem: first, a thread of uniform thickness, inside which polymers
are highly stretched by an extensional flow of constant extension
rate $\dot{\epsilon}$:
\beq
v_r(r,z,t) = -\dot{\epsilon}r/2, \quad
v_z(r,z,t) = \dot{\epsilon}z.
\label{ext}
\eeq
As a result, using the kinematic boundary condition at the interface,
the thread radius behaves as
\beq
h_{\rm thr} = h_0 e^{-\dot{\epsilon}t/2} \equiv h_0\ell,
\quad \ell = e^{-\dot{\epsilon}t/2},
\label{h_ext}
\eeq
where $h_0$ is a constant Second, fluid is injected from the thread into
a drop, in which elastic stresses have relaxed almost completely, and the
force balance is dominated by capillarity. Third, connecting the thread
and the drop is a corner region, the typical size of which is 
set by the thread thickness, for which we aim to find a
similarity description. A typical velocity scale is set by the
velocity $v_0 = \dot{\epsilon}L/2$ at the exit of the thread, based on
the linear axial velocity field $v_z$ (cf. \eqref{ext}), and a
vanishing velocity at the middle of the thread (on account of symmetry),
of total length $L$.

We begin by finding the solution inside the thread, where the radius
is assumed uniform, shrinking at an exponential rate, consistent with
the incompressible extensional velocity field (\ref{ext}). Owing to
translational invariance along the thread, we assume that the
polymeric stress is $z$-independent, but allow for an arbitrary
dependence in the radial direction. We expect stresses to be of the
same magnitude as the capillary pressure, which scales like
$\gamma / h_{\rm thr}$. However, since $h_0$ is still to be determined,
we now introduce the scaling
\beq
\sigma_{zz} = \frac{\gamma}{R_0\ell}\overline{\sigma}_0(\overline{r}) + C, \quad
\overline{r} = \frac{r}{R_0\ell}, 
\label{stress_zz_thread_a}
\eeq
where $\overline{\sigma}_0$ is the dimensionless axial stress 
inside the thread. Inserting \eqref{stress_zz_thread_a}
into (\ref{equ2}), the axial component of the terms on the left
are:
\[
\frac{\partial \sigma_{zz}}{\partial t} +
v_r\frac{\partial \sigma_{zz}}{\partial r} =
\frac{\dot{\epsilon}\gamma}{2 R_0 \ell}\overline{\sigma}_0 +
\frac{\dot{\epsilon}\gamma}{2 R_0 \ell}\overline{r}\overline{\sigma}_0' -
\frac{\dot{\epsilon}\gamma}{2 R_0 \ell}\overline{r}\overline{\sigma}_0' = 
\frac{\dot{\epsilon}\gamma}{2 R_0 \ell}\overline{\sigma}_0, 
\]
and thus the axial component of (\ref{equ2}) becomes
\[
\frac{\dot{\epsilon}\gamma}{2 R_0 \ell}\overline{\sigma}_0 = 
\left(2\dot{\epsilon} - \frac{1}{\lambda}\right)
\left(\frac{\gamma\overline{\sigma}_0}{R_0\ell} + C\right) +
\frac{2\dot{\epsilon}\eta_p}{\lambda}.
\]
For this to be a solution, we have
$\dot{\epsilon} = 2/(3\lambda)$ \cite[]{BER90,EH97} and
$C = -4\eta_p/\lambda$, so that the solution in the thread becomes
\beq
\sigma_{zz} = \frac{\gamma}{R_0\ell}\overline{\sigma}_0(\overline{r})
-\frac{4\eta_p}{\lambda}.
\label{stress_zz_thread}
\eeq
Here $\overline{\sigma}_0(\overline{r})$
inside the thread can have an arbitrary radial profile, whereas
usually a uniform profile has been assumed. The exact form of the
profile will be determined by the initial (non-universal) dynamics
of the thread. Analysing the other two components in a similar manner,
we find to leading order
\beq
\sigma_{rz} = \frac{\gamma}{R_0}\overline{\sigma}_{rz}^{(0)}
(\overline{r})\ell^2, \quad \overline{\sigma}_{rr} = -\frac{2\eta_p}{5\lambda}.
\eeq
Thus inside the thread, the axial stress $\sigma_{zz}$ dominates over the
remaining components.

\subsection{Similarity solution for the corner region}

We now describe the transition region at the end of the thread,
assuming that the thread thickness $h_{\rm thr}$ is the only length scale.
Again, the ratio $h_0/R_0$ is still to be determined, so for the similarity
description we use $\ell=\exp\{-t/3\lambda \}$ and the length $R_0$, so that
Thus using $\ell = \exp\{-t/(3\lambda)\}$, we find the similarity description
\beq
\label{sim}
h(z,t)= R_0\ell\overline{h}(\overline{z}), \quad
p(z,r,t) = \frac{\gamma}{R_0\ell} \overline{p}(\overline{z},\overline{r}), \quad
{\bf v}(z,r,t) = v_0\overline{\bf v}(\overline{z},
\overline{r}), \quad
\boldsymbol{\sigma}(z,r,t) =
\frac{\gamma}{R_0 \ell}\overline{\boldsymbol{\sigma}}
(\overline{z},\overline{r}),
\eeq
where $\overline{z} = z /(R_0 \ell)$ and $\overline{r} = r /(R_0 \ell)$.
Then to leading order as $\ell\rightarrow 0$, the similarity equations
become
\beqa
\label{sim_equ1}
&& \overline{\grad} \overline{p} =
\overline{\grad}\cdot\overline{\boldsymbol{\sigma}}_p
+ \overline{v}_0\overline{\triangle}\overline{\bf v},
\label{sim_equ2} \\
&& (\overline{\bf v}\cdot\overline{\grad})\overline{\boldsymbol{\sigma}}_{p} =
(\overline{\grad}\overline{\bf v})^T\cdot\overline{\boldsymbol{\sigma}}_p +
\overline{\boldsymbol{\sigma}}_p\cdot(\overline{\grad}\overline{\bf v})
\label{sim_equ3} \\
&& \overline{\grad}\cdot\overline{\bf v} = 0,
\eeqa
where $\overline{v}_0 = v_0\eta_s / \gamma =
\eta_s L / (3\lambda \gamma)$. The stress boundary condition is:
\beq
{\bf n}\cdot\left(\overline{\boldsymbol{\sigma}}_p
+ \overline{v}_0\overline{\dot{\boldsymbol{\gamma}}}\right) =
(\overline{p}-\overline{\kappa}){\bf n}.
\label{sim_stress_bc}
\eeq

\begin{figure}
\centering
\includegraphics[width=0.9\hsize]{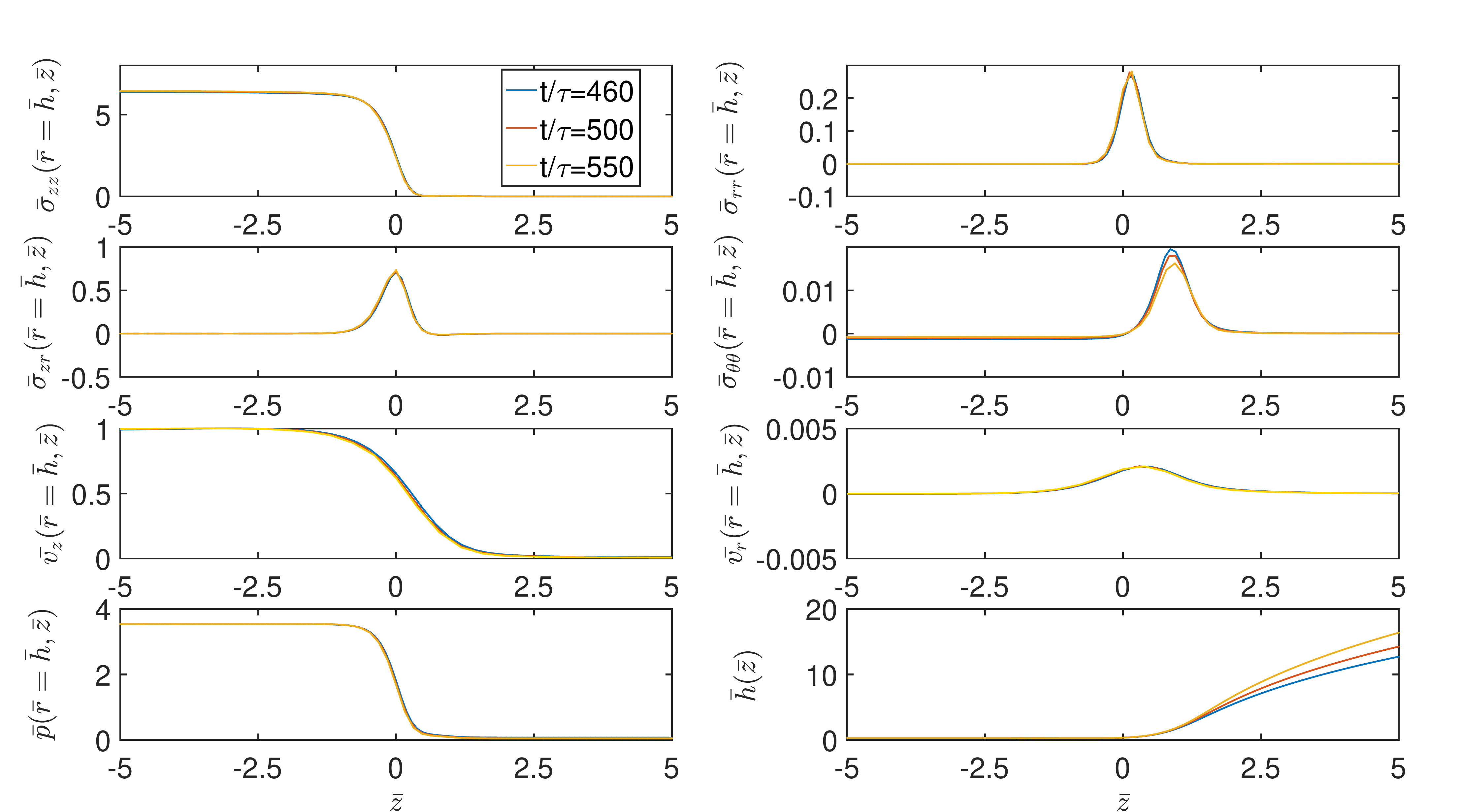}
\caption{\label{sim_prof}
Results of a numerical simulation with parameters as in
Fig.~\ref{bridge_exp}, rescaled to similarity variables according
to \eqref{sim}. The profiles for three different times collapse
onto a time-independent similarity solution. The top four profiles
are the components of the stress tensor, evaluated at the surface,
the bottom four profiles show the components of the velocity,
the pressure (again at the surface), as well as the surface profile
itself. All profiles collapse onto the time-independent similarity 
solution defined by \eqref{sim}. 
}
\end{figure}

\begin{figure}
\centering
\includegraphics[width=0.8\hsize]{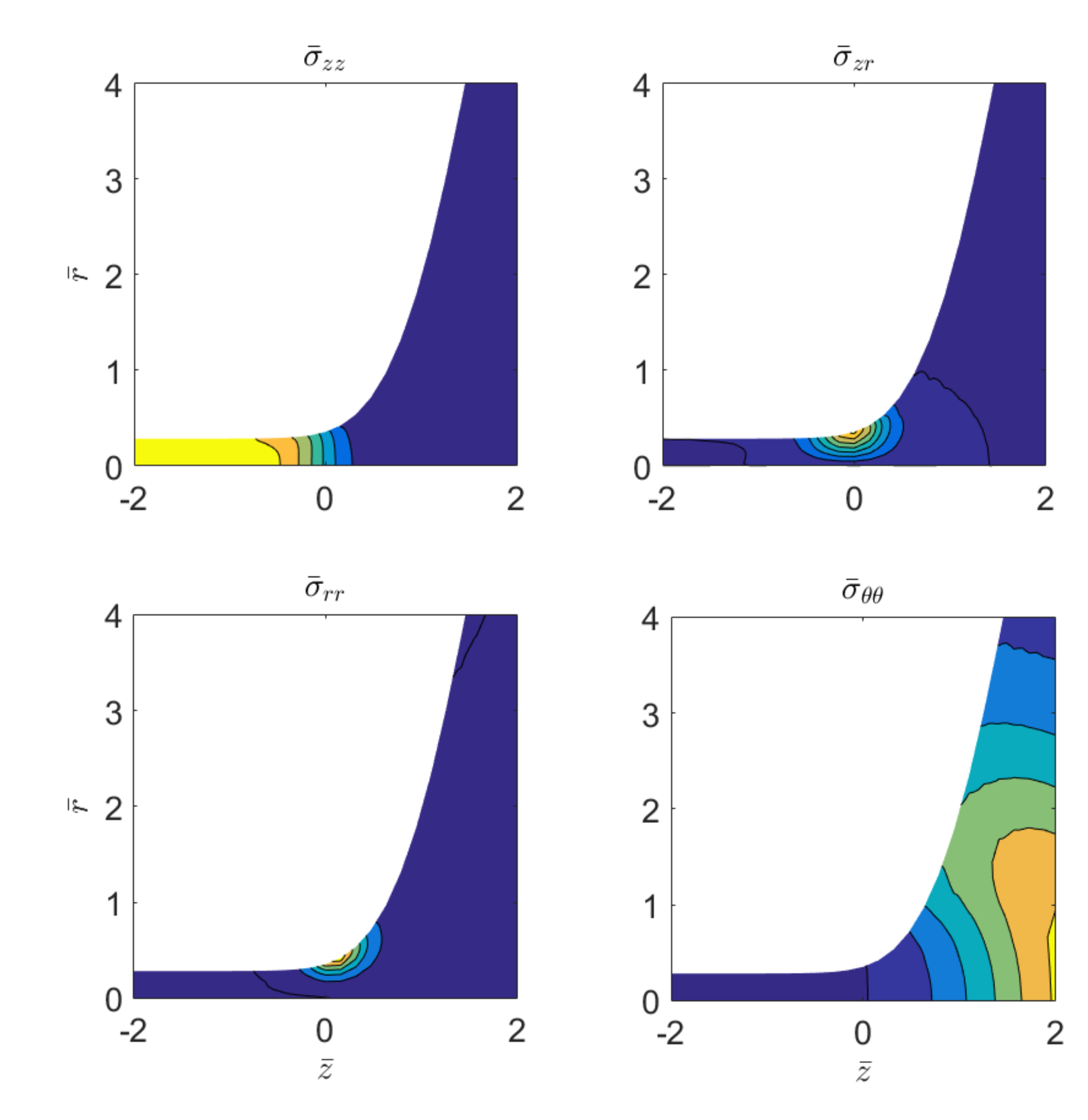}
\caption{\label{sim_stress}
Contour plots of stresses, rescaled according to \eqref{sim},
obtained as in Fig.~\ref{sim_prof}, for the last dimensionless
time $t / \tau = 550$.}
\end{figure}

The system (\ref{sim_equ1})-(\ref{sim_stress_bc}) does not contain time,
but there are matching conditions to be satisfied for
$\overline{z}\rightarrow\pm\infty$. For
$\overline{z}\rightarrow -\infty$ the solution has to tend toward the
thread solution
\beq
\overline{\bf v} \rightarrow {\bf e}_z, \quad
\overline{h} \rightarrow \overline{h}_0, \quad
\overline{p}\rightarrow 1/\overline{h}_0, \quad
\overline{\boldsymbol{\sigma}}_p\rightarrow \overline{\sigma}_0(\overline{r})
{\bf e}_z\otimes {\bf e}_z.
\label{bc_left}
\eeq
The first condition says that since the scale of the similarity
solution $h_{\rm thr}$ is much smaller than the length $L$ of the thread,
the velocity of the similarity solution has to match the velocity at
the exit of the thread. The other conditions correspond to thickness
and the stress distribution inside the thread. To verify the similarity
form \eqref{sim} we present our numerical results in scaled form in
Figures \ref{sim_prof} and \ref{sim_stress}. We indeed observe a
collapse of the data and moreover the numerics confirm the matching
conditions \eqref{bc_left}.

\begin{figure}
\centering
\includegraphics[width=\hsize]{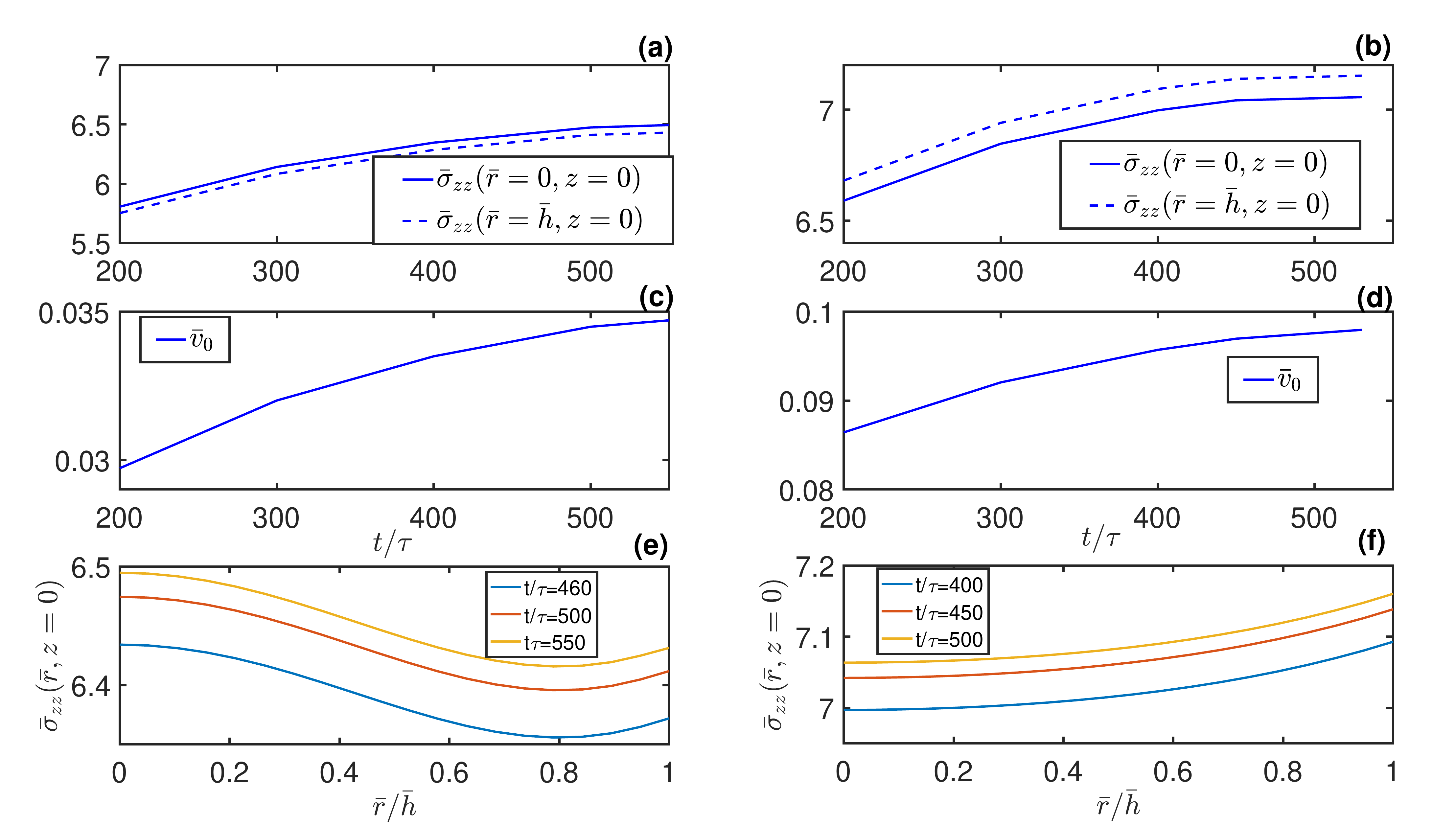}
\caption{\label{sim_v0}
A convergence study of the axial stress distribution inside the
uniform thread, taken at the middle of the thread $z=0$.
The left hand column (a), (c) and (e) corresponds to $S=\eta_s/\eta_0=0.25$
as in Fig.~\ref{bridge_exp}, the right hand column while (b),(d) and (f)
to $S=0.7$; all other parameters are as in Fig.~\ref{bridge_exp}. 
The top two panels demonstrate convergence toward a self-similar
stress distribution, which remains non-constant over the radius,
as the dimensionless exit velocity converges toward its asymptotic value
(two middle panels). The bottom two panels show the self-similar
stress profiles $\overline{\sigma}_0(\overline{r})$. The shape of
the profiles, scaled according to \eqref{sim}, depends on the solvent
ratio $S$. 
  }
\end{figure}

Our main aim will be to calculate the self-similar thread thickness
$\overline{h}_0$, which determines the prefactor in the exponential
thinning law \eqref{h_ext}. Clearly, $\overline{h}_0$ will depend
on the initial thread radius: the build-up of elastic tension 
depends on the history of deformation, and thus must involve the
ratio $h_{\rm thr}/R_0$. However, in obtaining the similarity
description \eqref{sim_equ1}-\eqref{sim_equ3}, the choice of the
length $R_0$ was somewhat arbitrary, in the sense that any fixed
length would give rise to the same similarity equations. The effect
of the initial jet radius will be properly introduced in
Sec.\ref{sec:elastic}, where $\overline{h}_0$ will be determined.
However, the product
$h \sigma_{zz}= \gamma \overline{h}_0 \overline{\sigma}_{zz}$ does
not depend on either $R_0$ or $\ell$, and is thus expected to be universal.
This means that with $\overline{h}_0$ in hand and measuring $h$,
one can deduce the extensional stress in the thread.

As implied by \eqref{stress_zz_thread} and \eqref{bc_left}, the
stress will in general not be constant across the fibre, corresponding
to a stress distribution $\overline{\sigma}_0(\overline{r})$ which depends on the
radius explicitly. This is confirmed in Fig.~\ref{sim_v0}, although
deviations from a constant stress are small, since the dimensionless
exit velocity $\overline{v}_0=0.035$ is small. Indeed, we will see below
that for $\overline{v}_0=0$ the radial stress distribution is
uniform. On the top we show the axial stress on the axis and on the
surface, which converge to different values in the long-time limit.
While for $S=0.25$ the stress is greater on the axis, for $S=0.7$ it
is the other way around.

On the bottom we show the corresponding profiles across the thread 
for the two different values of the solvent viscosity. The two
shapes are entirely different, and indicate a non-universal dependence
of the stress distribution on the initial dynamics of the collapsing
bridge. 

\subsection{Force balance}
Now we derive boundary conditions for $\overline{z}\rightarrow\infty$,
i.e. toward the drop. Owing to the scaling of the polymeric stress
$\sim 1/h  \sim e^{\dot{\epsilon}t/2}$, the total tension inside the
thread scales like $\sigma h^2 \sim \ell \sim e^{-\dot{\epsilon}t/2}$,
which needs to be supported by surface tension forces inside the drop
\cite[]{CEFLM06}. To derive this more formally, we follow \cite{EF05}
and integrate (\ref{sim_equ1}), written in the form
$\overline{\grad}\cdot\left(\overline{\boldsymbol{\sigma}}_p
+\overline{v}_0\overline{\dot{\boldsymbol{\gamma}}} - p{\bf I}\right)
\equiv\overline{\grad}\cdot\overline{\boldsymbol{\sigma}}= 0$,
over the fluid
volume bounded by the planes $\overline{z}=\overline{z}_{\pm}$.
Using the dynamic boundary conditions in the
form $\overline{\bf n}\cdot\overline{\boldsymbol{\sigma}} =
-\overline{\kappa}\overline{\bf n}$, we have
\beq
0 = \oint_S\overline{\bf n}\cdot\overline{\boldsymbol{\sigma}}ds =
-\int_O\overline{\kappa}\overline{\bf n} ds \pm\int_{C_{\pm}}
{\bf e}_z\cdot\overline{\boldsymbol{\sigma}},
\label{balance}
\eeq
where $S$ is the closed surface between the planes
$\overline{z}=\overline{z}_+$ and $\overline{z}=\overline{z}_-$,
denoted $C\pm$, as well as the jet surface $O$. According to \cite{EF05},
the integral over $O$ is
\[
\int_O\overline{\kappa}\overline{\bf n} ds =
-\left.\frac{2\pi\overline{h}}{\sqrt{1+\overline{h}_{\overline{z}}^2}}
\right|_{\overline{z}_-}^{\overline{z}_+}{\bf e}_z,
\]
and the integral over the $z$-component of $C_{+}$ and $C_{-}$ is
(using incompressibility to transform the viscous term)
\[
2\pi\left[\int_0^{\overline{h}}\overline{r}\overline{\sigma}^{(p)}_{zz}
  d\overline{r}
  - \int_0^{\overline{h}}\overline{r}\overline{p}d\overline{r} -
2\overline{v}_0\overline{h}\overline{v}_r(\overline{h},\overline{z})
  \right]_{\overline{z}_-}^{\overline{z}_+}. 
\]

We note further that
\[
\frac{2\overline{h}}{\sqrt{1+\overline{h}_{\overline{z}}^2}}-
2\int_0^{\overline{h}}\overline{r}\overline{\kappa}d\overline{r} =
\frac{2\overline{h}}{\sqrt{1+\overline{h}_{\overline{z}}^2}}
-\overline{h}^2\kappa = \overline{h}^2\overline{K}, 
\]
where
\beq
K = \frac{1}{h(1+h_{z}^2)^{1/2}} +
\frac{h_{zz}}{(1+h_{z}^{2})^{\frac{3}{2}}}, 
\label{K}
\eeq
so that \eqref{balance} can be written as
\beq
0 = \left.2\int_0^{\overline{h}}\overline{r}\overline{\sigma}^{(p)}_{zz}
d\overline{r} + 2\int_0^{\overline{h}}\overline{r}
(\overline{\kappa} - \overline{p})d\overline{r} -
4\overline{v}_0\overline{h}\overline{v}_r(\overline{h},\overline{z}) +
\overline{h}^2\overline{K}\right|_{\overline{z}_-}^{\overline{z}_+}.
\label{force_b}
\eeq
So far no approximation has been made.

Using the boundary conditions \eqref{bc_left} inside the thread,
\eqref{force_b} becomes
\beq
\overline{T} \equiv
\left.2\int_0^{\overline{h}_0}
\overline{r}\overline{\sigma}_0(\overline{r})d\overline{r}
+ \overline{h}_0 =
2\int_0^{\overline{h}}\overline{r}\overline{\sigma}^{(p)}_{zz}d\overline{r} 
+ 2\int_0^{\overline{h}}\overline{r}
(\overline{\kappa} - \overline{p})d\overline{r} -
4\overline{v}_0\overline{h}\overline{v}_r(\overline{h},\overline{z}) +
\overline{h}^2\overline{K}\right|_{\overline{z}=\overline{z}_+},
\label{force_b1}
\eeq
where $\pi \overline{T}$ is the constant tension in the liquid thread.
As we confirm below, all viscous and elastic stresses decay rapidly inside
the drop, and the balance is maintained by surface tension forces alone.
As a result, the deviatoric part of the stress $\overline{\sigma}_{zz}$
and $\overline{v}_r$ decay rapidly as $\overline{z}\rightarrow\infty$, 
and the pressure almost equals the capillary pressure:
$\overline{p}\approx\overline{\kappa}$. Only the fourth term on the right of 
\eqref{force_b1}, coming from surface tension, survives, and equation
describing the shape of the capillary region is \cite[]{EF_book}:
\beq
\overline{T} = \overline{h}^2\overline{K}. 
\label{capillary}
\eeq

This equation describes the crossover toward the drop, whose
dimensionless radius $\overline{R} = R/h_{\rm thr}$ diverges in the
limit. Indeed, in the intermediate region where
$\overline{h} \ll \overline{R}$, the only way the right hand side of
\eqref{capillary} can be constant is to require that
$\overline{h}/\overline{h}_{\overline{z}}\rightarrow const$, which
is achieved for
\beq
\overline{h} = He^{a\overline{z}}, \quad \overline{z}\rightarrow\infty. 
\label{growth}
\eeq
This means the right hand side of \eqref{capillary} becomes $2/a$
in the limit, and we obtain
\[
\overline{T} = \frac{2}{a}. 
\]

\section{The elastic correspondence}
\label{sec:elastic}
At least in the limit that the contribution from the solvent
in the momentum balance \eqref{sim_equ1} is negligible, one can
make use of the fact that in the similarity equations
\eqref{sim_equ1}-\eqref{sim_equ3} terms containing the relaxation
time $\lambda$ have disappeared. Thus, we anticipate that the same
set of equations can be obtained in a purely elastic description,
where we take the limit $\lambda \rightarrow \infty$ such that
elastic stress do not relax. 
In the absence of inertia the elastic equations describe the
equilibrium between surface tension, which tends to deform the elastic
medium, and elasticity, which resists deformation. Thus for
$\lambda \rightarrow \infty$, but keeping $\mu = \eta_p/\lambda$
fixed, this equilibrium state is described
by
\beqa
\label{elas_equ1}
&& \grad p =
\grad\cdot\boldsymbol{\sigma}_p, \\
\label{elas_equ2}
&& \overset{\triangledown}{\boldsymbol{\sigma}}_p =
\mu\dot{\boldsymbol{\gamma}}, \\
\label{elas_equ3}
&& \grad\cdot{\bf v} = 0,
\eeqa
using that $\partial_t\boldsymbol{\sigma}_p = 0$.
One can already see that in the limit of vanishing elasticity
$\mu \rightarrow 0$, \eqref{elas_equ1}-\eqref{elas_equ3} are identical
to the similarity equation \eqref{sim_equ1}-\eqref{sim_equ3} 
for the case where $v_0=0$ (the stress coming from the solvent
can be neglected). So indeed, the similarity solution can be computed
from the elastic limit. However, it is advantageous to first retain
the elastic term  $\mu \dot{\boldsymbol \gamma}$, as it will
allow to properly introduce the
history-dependence into the description, and to determine the value of
$\bar h_0$ we are looking for. 

To proceed, we first show (for details see \cite{ES_elastic}),
that instead of solving
\eqref{elas_equ2} with the constraint \eqref{elas_equ3}, one
can write $\boldsymbol{\sigma}_p$ as the stress tensor
of a neo-Hookean elastic solid, which depends quadratically on the
deformation from an unstressed reference state. Thus one
needs to find a transformation of the undeformed state (e.g.
a cylinder), parameterized by variables $R,Z$, into a deformed state
$r = r(R,Z), z = z(R,Z)$
such that elastic and surface tension forces
are balanced. The equilibrium state is characterised by the
deformation tensor (in Cartesians)
\beq
F_{iK} = \frac{\partial x_i}{\partial X_K},
\label{def}
\eeq
which satisfies incompressibility
\beq
\det{\bf F} =
\frac{r}{R}\left(\frac{\partial r}{\partial R}
\frac{\partial z}{\partial Z}-
\frac{\partial r}{\partial Z}
\frac{\partial z}{\partial R}\right)= 1.
\label{incompr}
\eeq
Here lower case variables and lower case indices refer to the deformed state,
upper case variables and upper case indices refer to the undeformed state.

If we write the stress tensor as
${\boldsymbol{\sigma}}_p = \mu\left({\bf A} - {\bf I}\right)$,
where ${\bf A}$ is the conformation tensor as in \eqref{equaNumeric3},
then \eqref{elas_equ2} simplifies to $\overset{\triangledown}{\bf A} = 0$.
Using the fact that ${\bf v} = d{\bf x}/dt$, where the time derivative
is taken at a constant material point ${\bf X}$, one shows that \cite[]{MS15}
\beq
\overset{\triangledown}{\bf A} = {\bf F}
\left[\frac{d}{dt}\left({\bf F}^{-1}{\bf A}{\bf F}^{-T}\right)\right]{\bf F}^T
\label{up_relation}
\eeq
From this relation, it is clear that ${\bf A}={\bf F} {\bf F}^T$
indeed provides an integral
to the equation $\overset{\triangledown}{\bf A} = 0$. In addition, this
solution satisfies the condition that $\boldsymbol{\sigma}_p=0$ in the
initial configuration, where ${\bf F}={\bf I}$. 

With the above steps we have managed to integrate \eqref{elas_equ2},
and have identified $\boldsymbol{\sigma}_p$, satisfying a stress-free
initial condition. Since $-\mu{\bf I}$ can be written as part of the
pressure $p$, the elastic stress tensor becomes
\beq
\boldsymbol{\sigma} = -p{\bf I} +
\boldsymbol{\sigma}_p
\equiv -p{\bf I} + \mu{\bf F}{\bf F}^T,
\label{stress}
\eeq
which is the classical result for the ``true'' stress tensor of a
new-Hookean solid \cite[]{Suo_notes_fin,Suo_notes_rubber}.
Then in terms of the deformation, the stress tensor becomes
\cite[]{Negahban12}
\beq
\boldsymbol{\sigma}_p =
\mu\begin{pmatrix}
\left(\frac{\partial r}{\partial R}\right)^2 +
\left(\frac{\partial r}{\partial Z}\right)^2
& 0 & \frac{\partial r}{\partial R}
\frac{\partial z}{\partial R}
+\frac{\partial r}{\partial Z}
\frac{\partial z}{\partial Z}\\
0 & \left(\frac{r}{R}\right)^2 & 0  \\
\frac{\partial r}{\partial R}
\frac{\partial z}{\partial R}
+\frac{\partial r}{\partial Z}
\frac{\partial z}{\partial Z}
& 0 & \left(\frac{\partial z}{\partial Z}\right)^2 +
\left(\frac{\partial z}{\partial R}\right)^2
\end{pmatrix} .
\label{true}
\eeq

The true elastic stress is formulated in the deformed state,
(just like the fluid stress tensor), and thus satisfies the
same equilibrium conditions as before, but with $v_s=0$:
\beq
\grad\cdot\boldsymbol{\sigma} = 0, \quad
\left.{\bf n}\cdot\boldsymbol{\sigma}_p\right|_S =
\gamma(p - \kappa){\bf n}.
\label{equ_elast}
\eeq
If $R_0$ is the radius of the unperturbed interface as usual,
the free surface $h(z)$ has the parametric representation
\beq
h(z(R_0,Z)) = r(R_0,Z).
\label{h_par}
\eeq
Transforming the equilibrium condition \eqref{equ_elast} to
Lagrangian coordinates, we obtain in cylindrical coordinates:
\beqa
&& \frac{\mu R}{r}\left[\frac{\partial^2 r}{\partial R^2} +
\frac{1}{R}\frac{\partial r}{\partial R} - \frac{r}{R^2} +
\frac{\partial^2 r}{\partial Z^2} \right] =
\frac{\partial p}{\partial R}\frac{\partial z}{\partial Z} -
\frac{\partial p}{\partial Z}\frac{\partial z}{\partial R},
\label{3d-axi1} \\
&& \frac{\mu R}{r}\left[\frac{\partial^2 z}{\partial R^2} +
\frac{1}{R}\frac{\partial z}{\partial R} +
\frac{\partial^2 z}{\partial Z^2} \right] =
\frac{\partial p}{\partial Z}\frac{\partial r}{\partial R} -
\frac{\partial p}{\partial R}\frac{\partial r}{\partial Z},
\label{3d-axi2}
\eeqa
for the radial and axial force balances, respectively. The
stress boundary conditions on the surface are
\beqa
\label{stress_n_el}
&&  -\left.
\frac{h_z^2\sigma_{zz}-2h_z\sigma_{rz}+\sigma_{rr}}{1+h_z^2}
=\gamma(\kappa - p)\right|_{r=h}, \\
\label{stress_t_el}
&& \left.h_z(\sigma_{rr}-\sigma_{zz})+(1-h_z^2)\sigma_{rz}
=0\right|_{r=h}.
\eeqa

\subsection{Elastic similarity equations}
We now analyse the elastic problem in the limit
$G = \mu R_0/\gamma \rightarrow 0$,
which corresponds to the similarity solution for the pinching of a
polymeric thread. We expect a static balance between elastic stresses
and surface tension to be established at a thread thickness which
scales like an elasto-capillary length scale $\ell_e$:
$z \propto r \propto \ell_e$. Here we have anticipated that as in the
similarity solution \eqref{sim}, both Eulerian coordinates scale in
the same way. By contrast, the Lagrangian coordinates do not have the
same scales for $R$ and $Z$. Namely, the radial coordinate $R$ must
be rescaled by the original radius $R_0$; on account of 
incompressibility \eqref{incompr} we have that
$Z \propto \ell_e^3 / R_0^2$. This implies that the $d/dZ \gg d/dR$, so
that the dominant stress contributions in \eqref{true} is of the order
$\boldsymbol{\sigma}_p \propto \mu R_0^4 / \ell_e^4$. The curvature $\kappa$
scales as the inverse radius $h^{-1} \propto \ell_e^{-1}$ of the thread
(cf. \eqref{kappa}),
and thus on account of the stress balance \eqref{equ_elast}
$\mu R_0^4 / \ell_e^4 \propto \gamma / \ell_e$. As a result, the
elasto-capillary length scale becomes
\beq
\ell_e = \left(\frac{\mu R_0^4}{\gamma}\right)^{1/3} = R_0 G^{1/3},
\label{le}
\eeq
which sets the thickness of the thread \cite[]{EF_book}, at which elastic
and capillary forces are balanced.

According to the scaling analysis above, we introduce the similarity
solution
\beq
z = \ell_e\phi(\overline{R},\overline{Z}), \quad
r = \ell_e\psi(\overline{R},\overline{Z}), \quad
p = (\gamma/\ell_e)\Pi(\overline{R},\overline{Z}),
\label{zr_ss}
\eeq
valid in the limit $G\rightarrow 0$, where $\overline{R} = R/R_0$
and $\overline{Z} = Z R_0^2/\ell_e^3$.
The rescaled thread thickness $\overline{h}$ is defined by
$h = G^{1/3}\overline{h}$, and $z = \ell_e\overline{z}$,
so the slope $h_z=\overline{h}_{\overline{z}}$ remains invariant
under the scalings, and
$h_{zz}=\ell^{-1}\overline{h}_{\overline{z}\overline{z}}$. As a result, the
curvature scales as $\kappa = \ell^{-1}\overline{\kappa}$, where
$\overline{\kappa}$ is defined as in \eqref{kappa}, but on the basis of
$\overline{h}(\overline{z})$. Note that with this scaling,
$\boldsymbol{\sigma}_p \propto \mu G^{-4/3}$, so that the right hand side
of \eqref{elas_equ2} becomes subdominant in the limit. As a result, the
elastic similarity equations defined by the above rescaling become
identical to \eqref{sim_equ1}-\eqref{sim_equ3} in the limit that
solvent effects are negligible ($v_0=0$), and the similarity profiles
(overlined variables defined above) are the similarity
profiles defined by \eqref{sim} in that limit. The only way $\mu$
still enters into the problem is through the initial condition,
for which we require that the conformation tensor ${\bf A} = {\bf I}$. 
The rescaling \eqref{sim}, on the other hand, still contains a free
length scale $R_0$, which we choose such that the similarity profiles
of both the time-dependent problem and of the elastic problem are the same. 

The key observation is that the ratio $G = (\ell_e/R_0)^3$ between a
characteristic axial scale and a radial scale in Lagrangian
coordinates becomes very small in the limit. As a result, radial
derivatives can be neglected relative to axial ones. It also means that
in the limit $G \rightarrow 0$ fluid elements become stretched out
in the axial direction, so that the self-similar region comes from
just a single point in the reference configuration. The radius $R_0$
has to be taken as the radius at that point in the reference configuration.

With the scalings \eqref{zr_ss}, the similarity equations become,
beginning with incompressibility,
\beq
\frac{\psi}{\overline{R}}\left(\psi_{\overline{R}}\phi_{\overline{Z}} -
\psi_{\overline{Z}}\phi_{\overline{R}}
\right) = 1. 
\label{inc_ss}
\eeq
The equilibrium conditions \eqref{3d-axi1} and \eqref{3d-axi2}
are, to leading order as $G \rightarrow 0$, 
\beq
\frac{\overline{R}}{\psi}\psi_{\overline{Z}\overline{Z}} =
\Pi_{\overline{R}}\phi_{\overline{Z}}
- \Pi_{\overline{Z}}\phi_{\overline{R}}, \quad
\frac{\overline{R}}{\psi}\phi_{\overline{Z}\overline{Z}} =
\Pi_{\overline{Z}}\psi_{\overline{R}}
- \Pi_{\overline{R}}\psi_{\overline{Z}}. 
\label{balance_ss}
\eeq
The key point is that the second radial derivatives on the right hand
side have disappeared, and as result the equations are of first order
only in the radial coordinates. 

The deviatoric stresses are at leading order
\beq
\sigma_{zz} = \frac{\gamma}{\ell_e}\phi_{\overline{Z}}^2,
\quad
\sigma_{rz} = \frac{\gamma}{\ell_e}\phi_{\overline{Z}}\psi_{\overline{Z}}, \quad
\sigma_{rr} = \frac{\gamma}{\ell_e}\psi_{\overline{Z}}^2,
\label{stress_ss}
\eeq
and so the boundary conditions \eqref{stress_n_el},\eqref{stress_t_el},
to be evaluated at the surface $\overline{R}=1$, become to leading order
\beqa
\label{stress_n_el_ss}
&&  -\frac{h_z^2\phi_{\overline{Z}}^2-2h_z\phi_{\overline{Z}}\psi_{\overline{Z}}
  + \psi_{\overline{Z}}^2}{1+h_z^2} =\overline{\kappa} - \Pi, \\
\label{stress_t_el_ss}
&& h_z\left(\psi_{\overline{Z}}^2-\phi_{\overline{Z}}^2\right)
+(1-h_z^2)\phi_{\overline{Z}}\psi_{\overline{Z}}=0.
\eeqa
But along the surface we have
\[
\left.\frac{\psi_{\overline{Z}}}{\phi_{\overline{Z}}}\right|_{R=1} =
\left.\frac{\partial r}{\partial z}\right|_{R=1} = h_z,
\]
and so dividing through \eqref{stress_n_el_ss} and
\eqref{stress_t_el_ss} by $\phi_{\overline{Z}}^2$, the left hand side of
both equations is seen to vanish. As a result, the tangential
stress balance \eqref{stress_t_el_ss} is satisfied identically
in the limit $G \rightarrow 0$, while the kinematic
boundary condition as well as the normal stress balance become
\beq
\label{bc_asymp}
\psi(1,\overline{Z}) = \overline{h}(\phi(1,\overline{Z})), \quad
\Pi(1,\overline{Z}) =\overline{\kappa}.
\eeq
Thus as \eqref{balance_ss} is only of first order in $\overline{R}$,
there is only one stress boundary condition to be satisfied instead of
two.

We now show that the pressure may be eliminated from the leading-order
equations. Multiplying the first equation of
\eqref{balance_ss} by $\phi_{\overline{Z}}$ and using \eqref{inc_ss},
we find
\[
\phi_{\overline{Z}}\phi_{\overline{Z}\overline{Z}} = \Pi_{\overline{Z}} +
\frac{\psi}{R}\psi_{\overline{Z}}\left(\Pi_{\overline{Z}}\phi_R -
\Pi_R\phi_{\overline{Z}}\right) = \Pi_{\overline{Z}} -
\psi_{\overline{Z}}\psi_{\overline{Z}\overline{Z}},
\]
using the second equation \eqref{balance_ss} in the second step.
This can be integrated over $\overline{Z}$ to
\beq
\frac{1}{2}\left(\phi_{\overline{Z}}^2 + \psi_{\overline{Z}}^2\right)
= \Pi + C,
\label{Bernoulli}
\eeq
where the constant of integration $C$ is in general still a function
of $\overline{R}$. In our particular case, we will see that in fact
$C=0$. The conservation law expressed by \eqref{Bernoulli} can be
shown to be a special case of conservation laws discovered by \cite{E75}.

\subsection{Numerical simulations of the elastic equations}
\label{sub:elas_num}
\begin{figure}
\centering
\includegraphics[width=0.8\hsize]{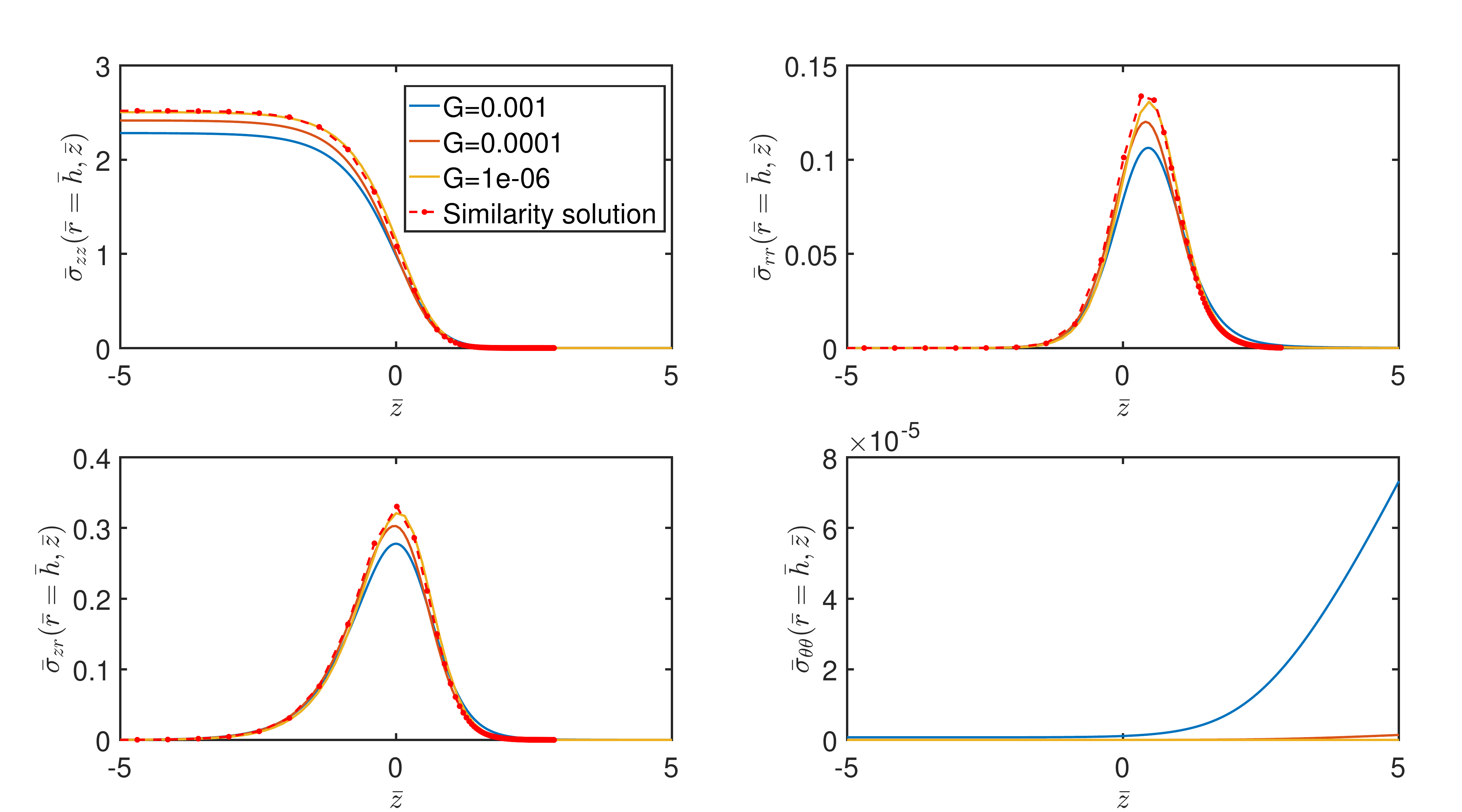}
\caption{\label{sim_elastic}
Convergence of the elastic solution toward the similarity solution
\eqref{zr_ss} for smaller and smaller values of the dimensionless
elasto-capillary number $G = \mu R_0/\gamma$. Stresses have been rescaled
according to \eqref{stress_rescale}. Profiles from the similarity solution as
described in Section~\ref{sub:sim_elas} are shown as the dashed line
for comparison. 
}
\end{figure}
Before we continue with solving the similarity equations, we test
the similarity ansatz \eqref{zr_ss} by rescaling full numerical simulations of
the elastic equations \eqref{stress}-\eqref{h_par}. 
For the numerical treatment, we define the reference state by
\[
R=R_0(Z)\eta, \quad Z=Z_0(\xi),
\]
with the elastic domain defined by $\eta \in [0,1]$ and $\xi \in [0,1]$
The function $Z_0(\xi)$ is designed to ensure a homogeneous
distribution of grid points in the deformed state, starting from a
uniform $\xi$-grid. This accounts for the extreme stretching in the
$Z$-direction for small values of $G$, and ensures that enough points
remain in the corner region.

As an initial condition, we take the free surface shape
\beq
h_{\rm in}(z) = 1 - \epsilon\cos(z/2),
\label{h0}
\eeq
where $\epsilon=0.005$. We are looking for two unknown
functions $f$ and $g$, where $r=r(R,Z)=f(\eta,\xi)$ and
$z=z(R,Z)=g(\eta,\xi)$, as well as the pressure $p(\eta,\xi)$.
These three unknowns are found from solving the three equations
\eqref{incompr} and \eqref{equ_elast}.
The free surface $h(z)$ then is given by the parametric representation
$h(g(1,\xi)) = f_1(1,\xi)$, from which the curvature $\kappa$ can be evaluated.

The  domain is discretized using fourth-order finite differences with
11 equally spaced points in the $\eta$-direction and 4001  points  in
the $\xi$-direction. The  resulting system of non linear equations
is solved using a Newton-Raphson  technique \cite[]{Herrada2016}. 
We start from the reference state as an initial guess
and $G$ sufficiently large ($G=100$) to ensure the convergence
of the  Newton-Raphson iterations. Once we get a solution, we use
this solution in a new run  with a smaller value of $G$.
As seen in Fig.~\ref{sim_elastic}, rescaled stresses as defined by
\beq
\overline{\boldsymbol{\sigma}} = \frac{\gamma}{\ell_e}\boldsymbol{\sigma}
\label{stress_rescale}
\eeq
converge nicely onto a universal similarity solution.

\subsection{The thread thickness}
\begin{figure}
\centering
\includegraphics[width=0.8\hsize]{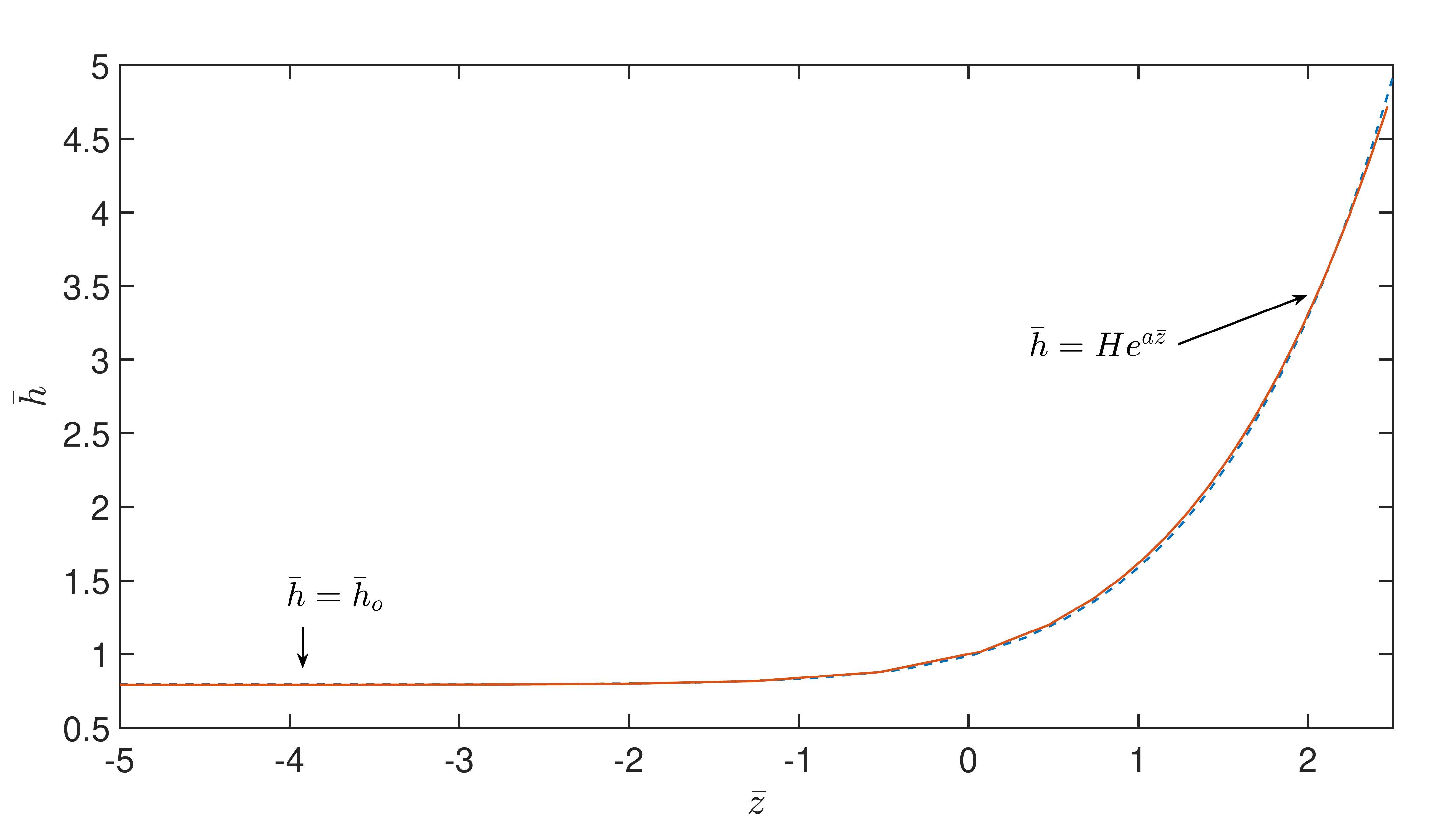}
\caption{The interface $\overline{h}(\overline{z})$ in
similarity variables. The solid line is the rescaled profile
obtained from the simulation for $G = 10^{-6}$, the dashed line
is the solution to the similarity equations. The thread thickness
is $\overline{h}_0$; for large arguments the profile grows exponentially.
}
\label{prof_fig}
\end{figure}
In solving the similarity equations, we first concentrate on the free
surface profile, shown in Figure~\ref{prof_fig} as a real-space profile
in similarity variables. We now show that the dimensionless
minimum thread thickness $\overline{h}_0$ can in fact be calculated
without first solving the full similarity equations. 
Inside the thread, a cylinder of constant dimensionless thickness
$\overline{R}=1$ (the reference state) is transformed into a thread
whose thickness $\overline{h}_0$ is constant. This results in the
transformation $\psi = \overline{h}_0 \overline{R}$, and thus from
incompressibility (cf. \eqref{inc_ss})
$\phi_{\overline{Z}} = 1/\overline{h}_0^2$. Then using \eqref{true}, 
the extensional part of the axial stress becomes
$\overline{\sigma}_0 = \phi_{\overline{Z}}^2 = 1/\overline{h}_0^4$.
In particular, the
radial stress profile $\overline{\sigma}_0(\overline{r})$ of
Section~\ref{sec:sim}
is uniform. Since in the elastic case there is no contribution from the
velocity, \eqref{force_b1} becomes
\beq
\overline{T} =
2\int_0^{\overline{h}} \overline{r}\overline{\sigma}_{zz}d\overline{r}
+ 2\int_0^{\overline{h}}\overline{r}
(\overline{\kappa} - \Pi)d\overline{r} +
\overline{h}^2\overline{K}.
\label{T_elast_ss}
\eeq
Evaluating \eqref{T_elast_ss} inside the thread of constant thickness,
one obtains a finite tension:
\beq
\overline{T} = \overline{h}_0^2/\overline{h}_0^4
+ \overline{h}_0 = \overline{h}_0 +\frac{1}{\overline{h}_0^2} > 0.
\label{tenion_finite}
\eeq
This corresponds to the earlier conclusion that there must be a positive
tension in a liquid bridge (polymeric or Newtonian) in order for
pinching to occur \cite[]{EF05,CEFLM06}.

On the other hand, toward the ``drop'' side of the elastic bridge,
where $\overline{h}$ becomes large, elastic stresses decay rapidly,
as is confirmed by our numerics. As a result, the balance \eqref{T_elast_ss}
is between $\overline{T}$ and the second term on the left, which must
approach a constant. As a result, it follows that for large $\overline{z}$,
the similarity profile must be of the form
\beq
\overline{h} = H e^{a\overline{z}}, \quad a = \frac{2}{\overline{T}} =
\frac{2\overline{h}_0^2}{1 + \overline{h}_0^3}.
\label{prof_lz}
\eeq
Using \eqref{prof_lz}, both radial and axial contributions to the
mean curvature \eqref{kappa} scale like $e^{-2a\overline{z}}$, and 
cancel to leading order, 
resulting in $\overline{\kappa} \sim e^{-4a\overline{z}}$.
The prefactor $H$ can be normalised to unity by shifting the coordinate
system. Thus putting $H=1$ is a way of fixing the origin, ensuring a
unique solution.

Applying the conservation law \eqref{Bernoulli} to the thread
surface $\overline{R}=1$, we have $\Pi = \overline{\kappa}$, and thus
\beq
\left.\frac{1}{2}\left(\phi_{\overline{Z}}^2 + \psi_{\overline{Z}}^2\right)
\right|_{\overline{R}=1} = \overline{\kappa} + C.
\label{surface_bal}
\eeq
The left hand side is the same as
${\rm tr}\left(\overline{\boldsymbol{\sigma}}\right)/2$, which vanishes
in the limit $\overline{z}\rightarrow\infty$. Since the curvature
vanishes as well for $\overline{z}\rightarrow\infty$, we have $C=0$.
On the other hand, for
$\overline{z}\rightarrow-\infty$, $\phi_{\overline{Z}} = 1/\overline{h}_0^2$ and
$\psi_{\overline{Z}} = 0$, while $\overline{\kappa} = 1 / \overline{h}_0$.
As a result, \eqref{surface_bal} yields
$1/(2 \overline{h}_0^4) = 1/\overline{h}_0$, and in summary
\beq
\overline{h}_0 = 2^{-1/3},  \quad \overline{\sigma}_0 = 2^{4/3},  \quad
\overline{T} = \frac{3}{2^{1/3}}, \quad
a = \frac{2^{4/3}}{3}.
\label{constants_sum}
\eeq
Thus we have calculated the dimensionless thread thickness without
explicitly solving the similarity equations.
Remarkably, these results are identical to those found using
lubrication theory \cite[]{EYa84,CEFLM06, EF_book}, although
the similarity solution doesn't satisfy the slenderness (small
slopes) requirement for lubrication to be valid. The reason is
that \eqref{constants_sum} only relies on the conservation laws
\eqref{surface_bal} and \eqref{T_elast_ss}, which are preserved in the
lubrication limit.

\subsection{The similarity solution}
\label{sub:sim_elas}
\begin{figure}
\centering
\includegraphics[width=0.8\hsize]{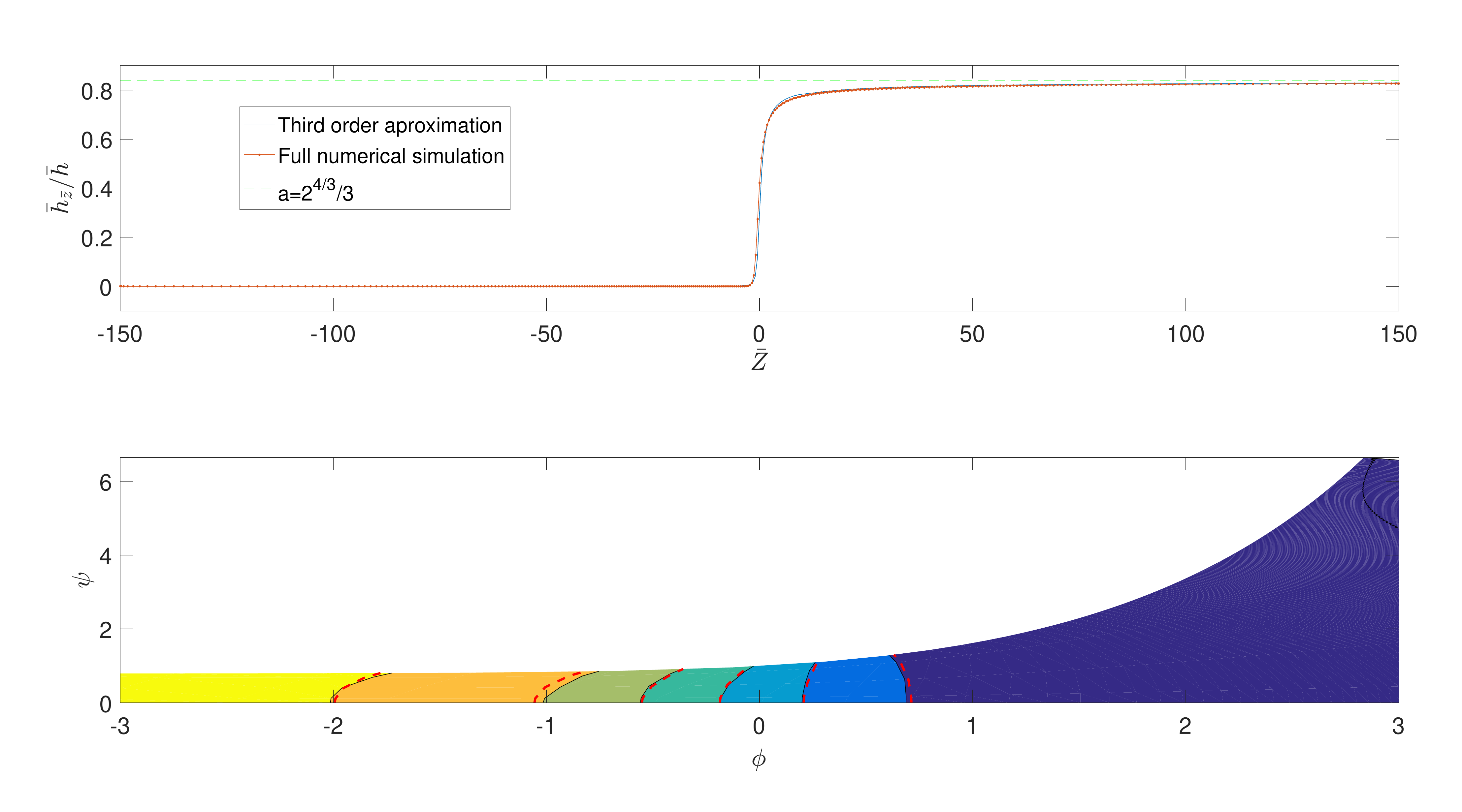}
\caption{\label{sim_profiles}
Convergence onto the similarity solution. On the top, the ratio
$\overline{h}_{\overline{z}} / \overline{h}$ approaches the limit
$a = 2^{4/3}/3$, over a wide range of $\overline{Z}$.
On the bottom, we show contours of the function $\Pi$ in the
$[\phi,\psi]$ domain, showing the transition region between the thread
and the drop region. Red dashes lines correspond with contours obtained
rescaling the pressure in the elastic simulation with $G=10^{-6}$.
Although the radial dependence is weak, the profile is not one-dimensional.
          }
\end{figure}
It remains to calculate the actual form of the profile and of the
deformation field inside. We can eliminate $\Pi$ from
\eqref{balance_ss}, which yields after simplification using \eqref{inc_ss}
\beq
\phi_{\overline{R}}\phi_{\overline{Z}\overline{Z}} +
\psi_{\overline{R}}\psi_{\overline{Z}\overline{Z}} =
\phi_{\overline{Z}}\phi_{\overline{Z}\overline{R}} +
\psi_{\overline{Z}}\psi_{\overline{Z}\overline{R}} .
\label{z-balance_ss}
\eeq
This can be solved together with the incompressibility constraint
\eqref{inc_ss}:
\[
\psi\left(\psi_{\overline{R}}\phi_{\overline{Z}} -
\psi_{\overline{Z}}\phi_{\overline{R}}
\right) = \overline{R}.
\]
On the surface $\overline{R}=1$, we have \eqref{bc_asymp}, which
in view of \eqref{surface_bal} takes the form
\beq
\label{bc_asymp_final}
\psi(1,\overline{Z}) = \overline{h}(\phi(1,\overline{Z})), \quad
\left.\frac{1}{2}\left(\phi_{\overline{Z}}^2 + \psi_{\overline{Z}}^2\right)
\right|_{R=1} = \overline{\kappa}.
\eeq
Differentiating the first equation of \eqref{bc_asymp_final}, it follows
at constant $\overline{R}=1$ that
$\psi_{\overline{Z}} = \phi_{\overline{Z}} \overline{h}_{\overline{z}}$,
and hence the second equation can also be rewritten
\beq
\frac{1}{2}\phi_{\overline{Z}}^2(1,\overline{Z}) = \frac{\overline{\kappa}}
{1 + \overline{h}_{\overline{z}}^2}.
\label{final2}
\eeq

For $\overline{Z}\rightarrow-\infty$ the boundary condition is
\beq
\psi = 2^{-1/3} \overline{R}, \quad \phi_{\overline{Z}} = 2^{2/3},
\label{as-}
\eeq
clearly there can be an arbitrary shift in $\overline{Z}$.
For $\overline{Z}\rightarrow\infty$ we must have
\beq
h = \psi(\overline{R}=1,\overline{Z}) = e^{a\phi(\overline{R}=1,\overline{Z})},
\label{as+}
\eeq
where the prefactor has been chosen $H=1$. For the other elastic fields
the boundary condition is that the pressure and all the elastic stresses
decay for $\overline{Z}\rightarrow\infty$.

To obtain the similarity solution numerically,  the domain
$[-L\leq \bar{Z}\leq L]\times [0\leq \bar{R}\leq 1]$ was discretized
using $n_{\bar{R}}$ and $n_{\bar{Z}}$ Chebyshev collocation points in the
$\bar{R}$ and $\bar{Z}$ directions, respectively. Here $L$ is the value
at which the $\bar{Z}$ coordinate is truncated. Since strong axial
gradients are expected near the origin $\bar{Z} = 0$ the grid is
stretched around that location using the stretching function
\beq
\bar{Z}=L \frac{\arctanh(\beta s)}{\arctanh(\beta)}. 
\eeq
Here $s$ is in the original Chebyshev interval $(-1\leq s\leq 1)$, and
$\beta$ a stretching parameter. 
 
Equations (\ref{inc_ss}) and (\ref{balance_ss}) with boundary conditions
given by  (\ref{bc_asymp_final}) and (\ref{as-}) are discretized in the
numerical domain. At $\bar{Z}=L$, the soft boundary  condition
$\phi_{\bar{Z}\bar{Z}}=\psi_{\bar{Z}\bar{Z}}=0$ is used. The resulting
discrete system of nonlinear equations have been solved using the
MATLAB function fsolve. As an initial guess for the nonlinear solver,
the numerical solution for the elastic simulation with $G=1 \times 10^{-6}$,
rescaled and interpolated into the similarity mesh was used.
Results presented have been obtained using $n_{\bar{R}}=5$, $n_{\bar{Z}}=450$ ,
$\beta=0.995$ and  $L=150$. A larger computational domain by imposing
a $L > 150$ does not alter the results. 

The resulting similarity profiles are shown in
Figs.~\ref{sim_elastic}-\ref{sim_profiles}. In Fig.~\ref{sim_elastic}
we compare representative stress profiles obtained from solutions
of the elastic problem with those calculated from the similarity
solution, in Fig.~\ref{prof_fig} the profile
$\overline{h}(\overline{z})$ is compared in greater detail. 
In Fig.~\ref{sim_profiles} (top) we show that the solution to the similarity
equations has converged in a large domain, using two different
methods. In particular, according to \eqref{growth} and \eqref{constants_sum}
$\overline{h}_{\overline z}$ converges to $a = 2^{4/3}/3$, as confirmed in 
Fig.~\ref{sim_profiles}.

Apart from a full numerical solution of the
two-dimensional similarity equations, we also show the result of a different
method of solution, explained in more detail in Appendix~\ref{app:radial}.
In view of the weak radial dependence, we expand the solution into a
power series \eqref{R_exp} in the radial direction. The results of
both methods of solution agree very well. 
On the bottom of Fig.~\ref{sim_profiles}, we show contours of the
self-similar pressure $\Pi$ in the $[\phi,\psi]$ domain. Once more,
agreement with contours obtained by rescaling the simulation for
$G = 10^{-6}$ agree very well with the full similarity solution. 

\begin{figure}
\centering
\includegraphics[width=\hsize]{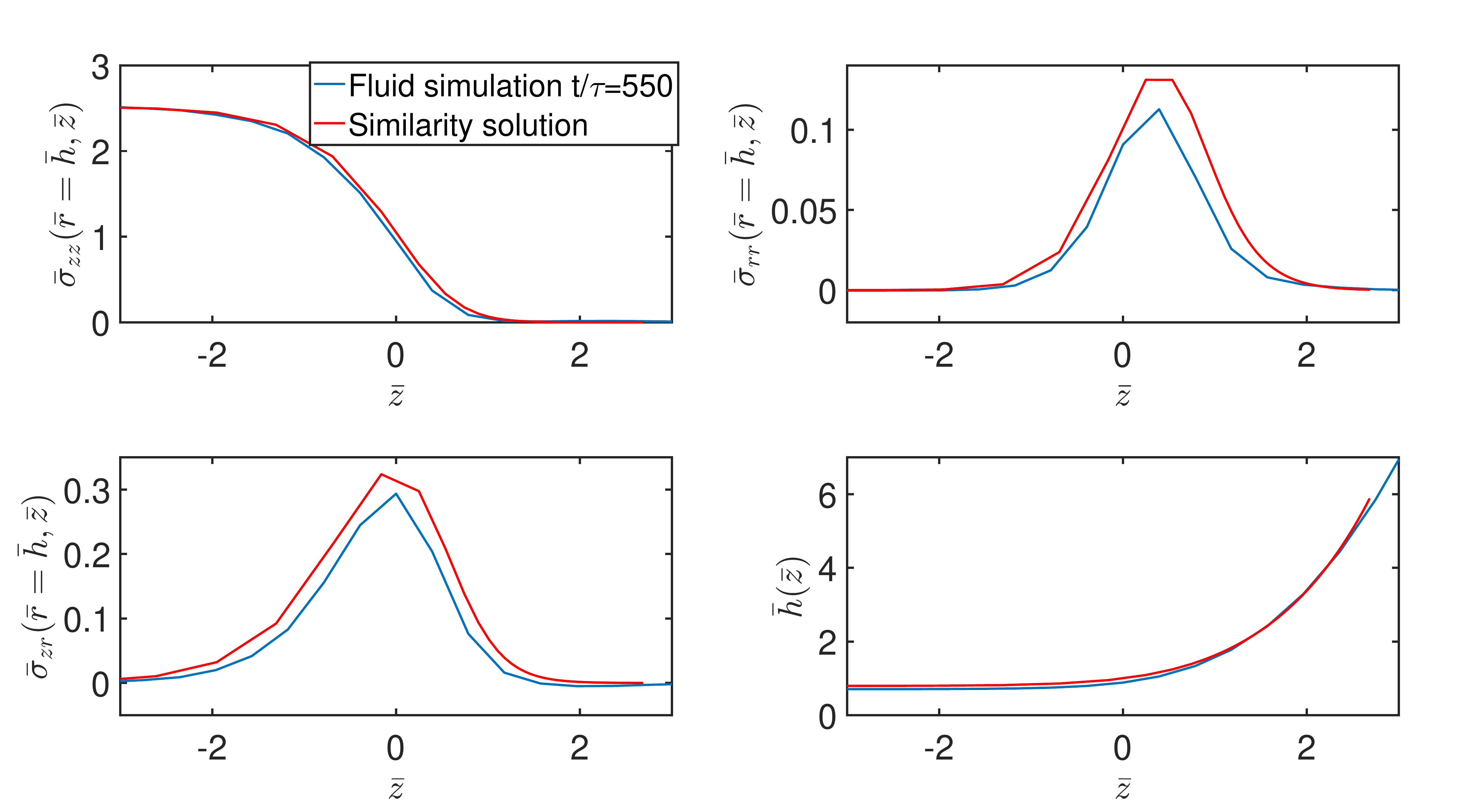}
\caption{\label{sim_viscoelastic}
Comparison of the similarity solution as described in
Section~\ref{sub:sim_elas} with the rescaled profiles \eqref{sim}
obtained from the time-dependent simulation shown in
Fig.\ref{bridge_exp} for ${\rm De}=60$. The time-dependent
simulation has been rescaled according to \eqref{sim}, with
an effective $R_0$ chosen such that
$\overline{\sigma}_{zz}=2^{4/3}$ inside the thread. 
  }
\end{figure}
As a final check, we return to the original problem of the
exponential thinning of a viscoelastic liquid bridge, as shown in
Fig.~\ref{sim_viscoelastic}. Since the dimensionless speed
$\overline{v}_0 = 0.035$ is small for our simulations, solvent
corrections are expected to be small. Indeed, there is good agreement 
between our similarity theory based on elastic effects alone, and
the full viscoelastic simulations. As we reiterate in the discussion
below, there is fading memory in the viscoelastic simulation, and
hence there is one adjustable parameter in the comparison, which we
choose so that $\overline{\sigma}_{zz} = 2^{4/3}$ inside the thread;
$\overline{h} = 2/\overline{\sigma}_{zz}$ is then predicted without
adjustable parameters. This is an effect of ${\rm De}$ being finite;
for the value of ${\rm De} = 60$ being used in the present simulation,
relaxation effects are almost negligible, as discussed below. 

\section{Discussion}
The aim of this paper has been to find similarity solutions describing the
time-dependent pinch-off of a polymeric thread. In the process, we found
that this similarity solution also describes another problem, the collapse
of a soft elastic bridge under surface tension. This fact had
been noticed before \cite{CEFLM06}, but is generalised here to the full
axisymmetric equations. In doing so, we make use of the very general
correspondence between fluid flow and non-linear elasticity
\cite[]{ES_elastic}.

A particular use of our theory is that we can now calculate
the radius of the thread that has formed in the middle of the collapsed
bridge. In the purely elastic case, returning to dimensional variables,
\eqref{constants_sum} implies that the thread thickness is
\beq
h_{\rm thr} = R_0\left(\frac{G}{2}\right)^{1/3},
\label{elas_thread}
\eeq
where $R_0$ is the initial radius of the elastic bridge. In the case of
very long relaxation times $\lambda$, such that an initial thread is
formed without the polymer having a chance to relax, followed by
exponential relaxation in the limit of long times, \eqref{elas_thread}
can be combined with \eqref{hmin} to give \cite[]{CEFLM06}
$h_0 = R_0 (G/2)^{1/3}$, and so
\beq
h_{\rm thr} = R_0\left(\frac{G}{2}\right)^{1/3}
\exp\left(-\frac{t}{3\lambda}\right), \quad {\rm De} \gg 1 .
\label{elas_thread_time}
\eeq
This is illustrated in the lower left panel of Fig.~\eqref{bridge_exp},
where the thread radius is shown for ${\rm De}=60$ and for
${\rm De}=\infty$, such that no relaxation takes place. The initial 
evolution is almost identical for the two case, and the thread radius
at the beginning
of the exponential relaxation (blue line) is just slightly below the
asymptotic value without any relaxation at all (red line). 

If however the Deborah number is intermediate, relaxation will
take place during formation of the thread, and the prefactor
$h_0$ will depend on the details of the initial dynamics.
However, if one measures the radius $h_{\rm thr}$ of the thread
during exponential thinning, the extensional stress can be deduced from
it using \eqref{constants_sum}:
\beq
\sigma_{zz} = \frac{2\gamma}{h_{\rm thr}}.
\label{h0_stress}
\eeq
This relation is needed for the original use of the liquid bridge geometry
as a rheometer \cite[]{BER90}: measuring the thread thickness, one can
measure the extensional viscosity $\eta_E = \sigma_{zz} / \dot{\epsilon}$
\cite[]{ABMK01}.

The present work provides one of the first examples of a similarity
solution in three dimensions, which does not reduce to an effective
one-dimensional theory \cite[]{CS97,DHL98,E_sing18}.
It is to be expected that non-trivial
similarity solutions in higher dimensions abound, but not many have
been found, owing to technical difficulties in obtaining them.
Another, and perhaps related feature concerns the universality of
similarity solutions. In a one-dimensional approximation, and taking
into account effects of the solvent viscosity, one obtains similarity
solutions which depend on the value of the parameter $\overline{v}_0$
alone \cite[]{CEFLM06}.

Here we provide evidence
(cf. Fig.~\ref{sim_v0}) that the three-dimensional, axisymmetric problem
is far less universal, and that there exists a similarity solution
for a given stress distribution $\overline{\sigma}_0(\overline{r})$ inside
the thread. To answer this question conclusively, one would have to
solve the full similarity equations \eqref{sim_equ1}-\eqref{sim_equ3}
for a prescribed profile $\overline{\sigma}_0(\overline{r})$, which we have not
yet attempted. It is however worth while keeping in mind that these
corrections coming from the solvent viscosity are small in practise.
For example, for the experiments by \cite{CEFLM06}, using a
quite elevated solvent viscosity, $\overline{v}_0 = 0.031$; for the
experiments by \cite{SWE08}, using water as a solvent,
$\overline{v}_0 = 1.1\times 10^{-3}$.

Finally, it is worth commenting on the experimental situation, 
since the Oldroyd B model considered here may be popular for its conceptual 
simplicity, but cannot be considered a first-principles description
of polymeric flow. While an exponential thinning regime of the polymer
thread is observed almost universally, indicating exponential stretching
of polymer strands, there are experimental indications of a more abrupt 
change in polymer conformation \cite[]{IK13}. As for the shape of the
self-similar profile, \cite{TLEAD18} find that the experimental free
surface profile grows significantly faster toward the drop when compared 
to full numerical simulations of the three-dimensional Oldroyd-B equations.
This means that the theoretical result $a = 2^{4/3}/3$ for 
the growth exponent in the free-surface profile \eqref{prof_lz}
underpredicts the experimentally observed value by about a factor of two. 
Now that a fully quantitative prediction of the Oldroyd-B fluid is finally
in place, and given the time that has passed since the original experiments
\cite[]{CEFLM06}, it would be well worth repeating the measurements
for a different system. Another avenue to pursue is to allow for polymer
models with more free parameters, such as a spectrum of relaxation
times. It might well be that different regions between the drop and the
thread are governed by different timescales.

\section*{Acknowledgments}
We gratefully acknowledge illuminating discussions with Marco Fontelos
during the early stages of this research. 
J. E. acknowledges the support of Leverhulme Trust International Academic
Fellowship IAF-2017-010, and is grateful to Howard Stone and his group
for welcoming him during the academic year 2017-2018. 
M. A. H. thanks the Ministerio de Econom\'{\i}a y competitividad for
partial support under the Project No. DPI2016-78887-C3-1-R
J.H.S. acknowledges support from NWO through VICI Grant No. 680-47-632

\begin{appendix}
\section{Force balance in Lagrangian coordinates}
\label{app:force}
In order to solve the similarity equations by an expansion in the radial
variable, described in Appendix~\ref{app:radial} below, it is useful to
implement the force balance \eqref{T_elast_ss} in Lagrangian coordinates.
To this end, within the leading order balance \eqref{stress_ss}, we use
the $z$-component of \eqref{balance}
\beq
\int_O\overline{\kappa}{\bf n} ds =
-\left.\frac{2\pi \overline{h}}{\sqrt{1+\overline{h}_{\overline{z}}^2}}
\right|_{\overline{Z}_-}^{\overline{Z}_+} =
\int_{C_{\pm}}{\bf n}_{\pm}\cdot\overline{\boldsymbol{\sigma}}\cdot{\bf e}_zds. 
\label{balance_z}
\eeq
However now $O$ is the surface of the thread between two fixed Lagrangian
coordinates $\overline{Z}_{\pm}$, and $C_{\pm}$ are the surfaces
\beq
\overline{z} = \phi(\overline{R},\overline{Z}_{\pm}),\quad
\overline{r} = \psi(\overline{R},\overline{Z}_{\pm}),
\label{L_surface}
\eeq
parameterized by $\overline{R}$ between 0 and 1; here ${\bf n}_{\pm}$ are the
outward normals to these surfaces. Inside the thread, $\overline{z}$
is constant, and hence ${\bf n}_- = -{\bf e}_z$, while on the right
\[
{\bf n}_+ = (\psi_{\overline{R}},-\phi_{\overline{R}})/
\sqrt{\phi_{\overline{R}}^2+\psi_{\overline{R}}^2},
\]
which follows from the parameterization \eqref{L_surface}.

The integral over the deviatoric part of the stress is
\[
\begin{split}
\int_{C_+}{\bf n}_+\cdot\overline{\boldsymbol{\sigma}}_p\cdot{\bf e}_zds &= 
2\pi\int_0^1{\bf n}_+\cdot\overline{\boldsymbol{\sigma}}_p\cdot{\bf e}_z
\psi \sqrt{\phi_{\overline{R}}^2+\psi_{\overline{R}}^2} d\overline{R} = \\ &=
2\pi\int_0^1\left(\psi_{\overline{R}}\phi_{\overline{Z}}^2 -
\phi_{\overline{R}}\psi_{\overline{Z}}\phi_{\overline{Z}}\right)\psi d\overline{R} =
2\pi\int_0^1 \phi_{\overline{Z}} \overline{R} d\overline{R},
\end{split}
\]
having used incompressibility \eqref{inc_ss} in the last step.
The pressure contribution is
\[
-\int_{C_+}\Pi{\bf n}_+\cdot{\bf e}_zds =
-2\pi\int_0^1\Pi\psi_{\overline{R}}\psi d\overline{R}.
\]
Inside the thread,
\[
{\bf n}_-\cdot\overline{\boldsymbol{\sigma}}\cdot{\bf e}_z =
-\boldsymbol{\sigma}_{zz} = \frac{1}{2}\left(\psi_{\overline{Z}}^2 + 
\phi_{\overline{Z}}^2\right) = -2^{1/3},
\]
having used \eqref{as-} and \eqref{constants_sum}. Thus the integral over
the cross section $\pi \overline{h}_0^2$ yields
\[
\int_{C_-}{\bf n}_{\pm}\cdot\overline{\boldsymbol{\sigma}}\cdot{\bf e}_zds
 = -\pi 2^{-1/3}.
\]
Finally, since inside the thread $\overline{h} = \overline{h}_0$,
\eqref{balance_z}
yields
\beq
\frac{2 \overline{h}}{\sqrt{1+\overline{h}_{\overline{z}}^2}} +
2\int_0^1 \phi_{\overline{Z}} \overline{R} d\overline{R} -
\int_0^1\left(\phi_{\overline{Z}}^2 + \psi_{\overline{Z}}^2\right)
\psi_{\overline{R}}\psi d\overline{R}
= \overline{T} = \frac{3}{2^{1/3}},
\label{balance_z_final}
\eeq
where we inserted the expression \eqref{Bernoulli} for the pressure,
with $C=0$.

\section{Solution by radial expansion}
\label{app:radial}
The similarity profiles only have a weak dependence in the
$\overline{R}$-direction, and are thus well represented by a polynomial
in $\overline{R}$. Thus we seek a solution by expanding in the
$\overline{R}$-coordinate according to
\beq
\phi = \phi_0 + \phi_2 \overline{R}^2 + \dots, \quad
\psi = \psi_1\overline{R} + \psi_3 \overline{R}^3 + \dots.
\label{R_exp}
\eeq
First, from incompressibility we have
\beq
\psi_1 = \frac{1}{\phi_0'^{1/2}}, \quad
\psi_3 = -\frac{\phi_0'\phi_2' + \phi_0''\phi_2}{4\phi_0'^{5/2}}.
\label{R_inc}
\eeq
Expanding \eqref{z-balance_ss} in a power series in $\overline{R}$ we
find at first order that
\beq
-4\phi_0'^4\phi_2' + 4\phi_0'^3\phi_0''\phi_2 - \phi_0'\phi_0'''
+ \phi_0''^2 = 0,
\label{R_1st}
\eeq
which is valid exactly, as an equation between Taylor coefficients.

The second equation is \eqref{final2}, which can be satisfied
approximately by truncating the expansion \eqref{R_exp} after the
first two terms. Hence we obtain
\beq
\frac{\left(\phi_0'+\phi_2'\right)^2}{2} = \frac{\overline{\kappa}}
{1 + \overline{h}_{\overline{z}}^2},
\label{final2_R}
\eeq
where
\[
\overline{h} = \psi_1 + \psi_3, \quad
\overline{h}_{\overline{z}} = \frac{\psi_1' + \psi_3'}{\phi_0' + \phi_2'}
\quad
\overline{h}_{\overline{z}\overline{z}} =
\frac{\psi_1'' + \psi_3''}{\left(\phi_0' + \phi_2'\right)^2} -
\frac{\left(\psi_1' + \psi_3'\right)\left(\phi_0'' + \phi_2''\right)}
{\left(\phi_0' + \phi_2'\right)^3},
\]
and $\psi_1,\psi_3$ is expressible through $\phi_0$ and $\phi_2$
using \eqref{R_inc}.
We have to solve the system of equations \eqref{R_1st},\eqref{final2_R}
for the two variables $\phi_0$ and $\phi_2$ with conditions
$\phi_0' = 2^{2/3}$ and $\phi_2 = 0$ for $\overline{Z}\rightarrow-\infty$.

A problem with \eqref{final2_R} is that it is of quite
high order. A much better alternative is to use
\eqref{balance_z_final}, because it only requires first derivatives
$\overline{h}_{\overline{z}}$, and because it implements the force balance
exactly; inserting \eqref{R_exp}, the integrals can be done easily.
The approximated solution was obtained  with the same numerical procedure
used to obtain the solution to the full similarity equation.
\end{appendix}

\bibliographystyle{jfm}

\end{document}